\documentclass[a4paper,fleqn]{cas-dc}

\usepackage[numbers]{natbib}
\usepackage{graphicx}
\usepackage{multirow}
\usepackage{threeparttable}
\usepackage{booktabs}
\usepackage{balance}
\usepackage{amsfonts}
\usepackage{amssymb}
\usepackage{algorithm}
\usepackage{algorithmic}
\usepackage{array}
\usepackage{multirow}
\usepackage{enumitem}

\usepackage{threeparttable}
\usepackage{balance}

\def\tsc#1{\csdef{#1}{\textsc{\lowercase{#1}}\xspace}}
\tsc{WGM}
\tsc{QE}

\begin{document}
\let\WriteBookmarks\relax
\def\floatpagepagefraction{1}
\def\textpagefraction{.001}
\let\printorcid\relax 

\shorttitle{A Survey on Fairness-aware Recommender Systems}    

\shortauthors{Jin et al.}

\title[mode = title]{A Survey on Fairness-aware Recommender Systems}  

\tnotemark[1]

\tnotetext[1]{This project is funded by the National Science Foundation (No: 62272340) and ARC Future Fellowship (No: FT210100097).}

\author[1]{Di Jin}
\ead{jindi@tju.edu.cn} 

\author[1]{Luzhi Wang}
\ead{wangluzhi@tju.edu.cn}

\author[2]{He Zhang}
\ead{he.zhang1@monash.edu}

\author[2]{Yizhen Zheng}
\ead{yizhen.zheng1@monash.edu} 

\author[3]{Weiping Ding}
\ead{dwp9988@163.com}

\author[4]{Feng Xia}
\ead{f.xia@ieee.org} 

\author[5]{Shirui Pan}
\ead{s.pan@griffith.edu.au} 
\cormark[1] 

\address[1]{College of Intelligence and Computing, Tianjin University, China}
\address[2]{Department of Data Science and AI, Faculty of IT, Monash University, Australia}
\address[3]{School of Information Science and Technology, Nantong University, China}
\address[4]{School of Computing Technologies, RMIT University, Australia}
\address[5]{School of Information and Communication Technology, Griffith University, Australia}

\cortext[1]{Corresponding author} 

\begin{abstract}
As information filtering services, recommender systems have extremely enriched our daily life by providing personalized suggestions and facilitating people in decision-making, which makes them vital and indispensable to human society in the information era. However, as people become more dependent on them, recent studies show that recommender systems potentially own unintentional impacts on society and individuals because of their unfairness (e.g., gender discrimination in job recommendations). To develop trustworthy services, it is crucial to devise fairness-aware recommender systems that can mitigate these bias issues.
In this survey, we summarise existing methodologies and practices of fairness in recommender systems. Firstly, we present concepts of fairness in different recommendation scenarios, comprehensively categorize current advances, and introduce typical methods to promote fairness in different stages of recommender systems. Next, 
after introducing datasets and evaluation metrics applied to assess the fairness of recommender systems, we will delve into the significant influence that fairness-aware recommender systems exert on real-world industrial applications.
Subsequently, we highlight the connection between fairness and other principles of trustworthy recommender systems, aiming to consider trustworthiness principles holistically while advocating for fairness. Finally, we summarize this review, spotlighting promising opportunities in comprehending concepts, frameworks, the balance between accuracy and fairness, and the ties with trustworthiness, with the ultimate goal of fostering the development of fairness-aware recommender systems.

\end{abstract}




\maketitle

\section{Introduction}
Recommendation systems (RSs) are information filtering systems that are expected to suggest products and services, i.e., items, that most likely interest a user \cite{ricci2022recommender}. The suggestions are related to various decision-making processes for a user, such as which products to purchase, which videos to watch and which songs to listen. As the world becomes more information-overloaded, recommender systems are particularly useful when users need to choose an item from an overwhelming selection offered by a service. Recommender systems are pervasive and have been utilised in various fields including e-commerce \cite{kersbergen2021learnings,gu2021self,guo2022intelligent}, economics \cite{yang2018novel,toquica2022recommender, bogaert2019evaluating}, education \cite{hassan2021learning,chuang2021moocers,agarwal2022knowledge}, etc. For the e-commerce industry, Amazon, for example, distributes product information provided by merchants to users based on their history of purchases or website interaction. Remarkably, its recommendation engine accounts for three-quarters of its total revenue \cite{qin2020attribute}. In the economics industry, the loans domain can be recommended to users based on their attributes information such as annual income, address, occupation, and so on \cite{musto2021fairness}. Lastly, in the education industry, Massive Open Online Course (MOOC), one of the leading online learning providers, recommends courses to users based on their historical opinions \cite{Mooc}. 

However, as the development of recommender systems surges and the reliance on them grows, these systems can lead to huge negative impacts on society and individuals due to unfairness. The causes of unfairness and their undesirable outcomes are numerous and significant. For instance, a loan domain recommender system may recommend loan domains influenced by a user's sensitive personal attributes, such as age, gender, and race.
As a result, minority groups such as senior citizens, women workers, and ethnic minorities may suffer worsening financial situations by being recommended loan domains with higher interest rates, which is obviously unfair. The second example is MOOC course recommendations \cite{Mooc}, in which more than $40\%$ of the courses are taught by American teachers, while the remaining courses from $73$ countries are rarely recommended to or enrolled in by online users. Due to the vast geographic imbalance in MOOC recommendations, teachers in smaller or less-known places faced a great disadvantage when attracting students. In a similar way, Amazon tended to recommend items from larger merchants over smaller merchants. Because of this, smaller merchants would find it difficult to compete with large merchants, even if they offered better prices or quality. Although recommender systems are supposed to mitigate these unfairnesses, they can actually exacerbate them in the recommendation pipeline.

Driven by these unintentional issues, people increasingly yearn for fairness since it is critical and essential in developing recommender systems that can be trusted. 
First, fairness benefits users, items, and even recommender systems themselves \cite{10.1145/3547333}. For example, in a fair recommender system, users can obtain more relevant information, including niche information, which can aid in breaking out of the cocoon of information. As minor items are allocated more exposure, the Matthew effect \cite{DBLP:conf/www/LiCFGZ21} is lessened, encouraging providers to enhance their creativity and diversity. 
By providing equal quality of service for objects from diverse backgrounds, fair recommender systems can also gain long-term interest due to positive feedback from users and item providers.
Second, a global consensus has recently been built on enhancing the trustworthiness of AI systems \cite{DBLP:journals/corr/abs-2107-06641, zhang2022trustworthy, DBLP:journals/corr/abs-2209-10117}, including fairness, robustness, explainability, privacy, etc. Devising fairness-aware recommender systems directly contribute to the trustworthiness of RSs \cite{DBLP:journals/corr/abs-2209-10117}. Moreover, studying the connections between fairness and other aspects (e.g., robustness \cite{DBLP:conf/sacmat/Fang0MS22}) benefits the comprehensive building of trustworthy systems \cite{DBLP:journals/corr/abs-2107-06641, zhang2022trustworthy}.
Finally, compliance with laws and regulations \cite{10.1145/3547333} requires recommender systems to be fair when interacting with people because fairness is one of the cornerstones of keeping social order. For example, discrimination against vulnerable groups of people based on sensitive information (e.g., gender, age, race \cite{lesota2021analyzing,wu2021learning}) is forbidden by current anti-discrimination laws \cite{holmes2005anti}, which also require that similar people should be treated similarly to ensure equality.

In this survey, \textbf{fair recommender systems} refer to recommender systems that can adapt to different users/items and provide indiscriminate recommendation services to them. Due to the diversity of scenarios, 
the concept of fairness within recommender systems can be interpreted in several unique ways, including group or individual fairness, static or dynamic fairness \cite{DBLP:journals/tois/MansouryAPMB22, ge2021towards}, single-sided or multi-sided fairness, etc. In addition to general metrics, 
task-specific special metrics (e.g., diversity, serendipity), can also be used to evaluate the fairness of recommender systems.
Although the fairness of general machine learning tasks (e.g., image classification) has been extensively explored \cite{DBLP:journals/csur/MehrabiMSLG21}, several unique characteristics of recommender systems differentiate their fairness study from that of general classification tasks. 
First, the fairness concerning multi-party benefits (e.g., multi-sided fairness) or dynamic evolution (e.g., dynamic fairness \cite{ge2021towards}) should be taken into account since recommender systems interact with both users and item providers in a dynamic process. 
Second, unfairness exists in the whole life-cycle of recommender systems because of their dynamic interactions with users and items (as shown in Fig. \ref{bias}). This fact requires researchers and practitioners to enhance fairness at each stage of developing recommender systems to avoid learning historical unfairness from the last loop and even amplifying it into the subsequent feedback loop. 
Therefore, it is imperative to summarise current efforts and advancements in building fairness-aware recommender systems, which contribute to developing trustworthy, responsible, and socially beneficial AI services.

Related reviews include surveys on recommender systems, fair recommender systems, and trustworthy recommender systems. 
Our survey differs from those surveys in that it elaborates on the existing advancements in the fair recommender systems as well as how fairness interacts with other aspects of trustworthy recommender systems. 
\textbf{(1)} Surveys on recommender systems review the advancements of performance-oriented methods on recommender systems. Wu et al. \cite{wu2020graph} explore the impact of Graph Neural Networks on different categories of recommender systems. Deldjoo et al. \cite{DBLP:journals/csur/DeldjooNM21} investigate the impact of adversarial learning on the security and accuracy of recommender systems. 
Wu et al. \cite{wu2022survey} introduce the modeling of collaborative filtering techniques in different types of recommender systems (e.g., content-rich recommender systems, sequential recommender systems). 
\textbf{(2)} Recently, surveys on fair recommendation systems have also emerged. For example, Wang et al. \cite{10.1145/3547333} 
explore various perspectives on unfairness issues and provide an overview of existing approaches, which include data-oriented, ranking, and re-ranking methods. 
Zehlike et al. \cite{zehlike2022fairness} delve into how ranking methods impact the fairness of recommender systems. They introduce the fair ranking framework and summarize the evaluation of fair ranking methods. Chen et al. \cite{DBLP:journals/corr/abs-2010-03240} describe the existence of biases in recommender systems and describe ways to address them.
Unlike these surveys, our survey categorizes current methods into pre-processing, in-processing, and post-processing methods; we also present comprehensive and fine-grained taxonomy on each of them (e.g., data re-labeling, data re-sampling, and data modification in pre-processing methods).
\textbf{(3)} Current surveys on trustworthy recommender systems allocate their attention to several different aspects of trustworthiness (e.g., explainability, robustness) \cite{DBLP:journals/corr/abs-2207-12515, DBLP:journals/corr/abs-2209-10117} or present a conceptual framework of trustworthy recommender systems \cite{DBLP:journals/corr/abs-2208-06265}. In contrast, our survey focuses on comprehensively summarising current advancements in fairness and discussing the influence on other aspects of trustworthy recommender systems from the view of enhancing fairness.

\begin{figure*}[htpb]
	\centering
	\includegraphics[width=0.95\textwidth]{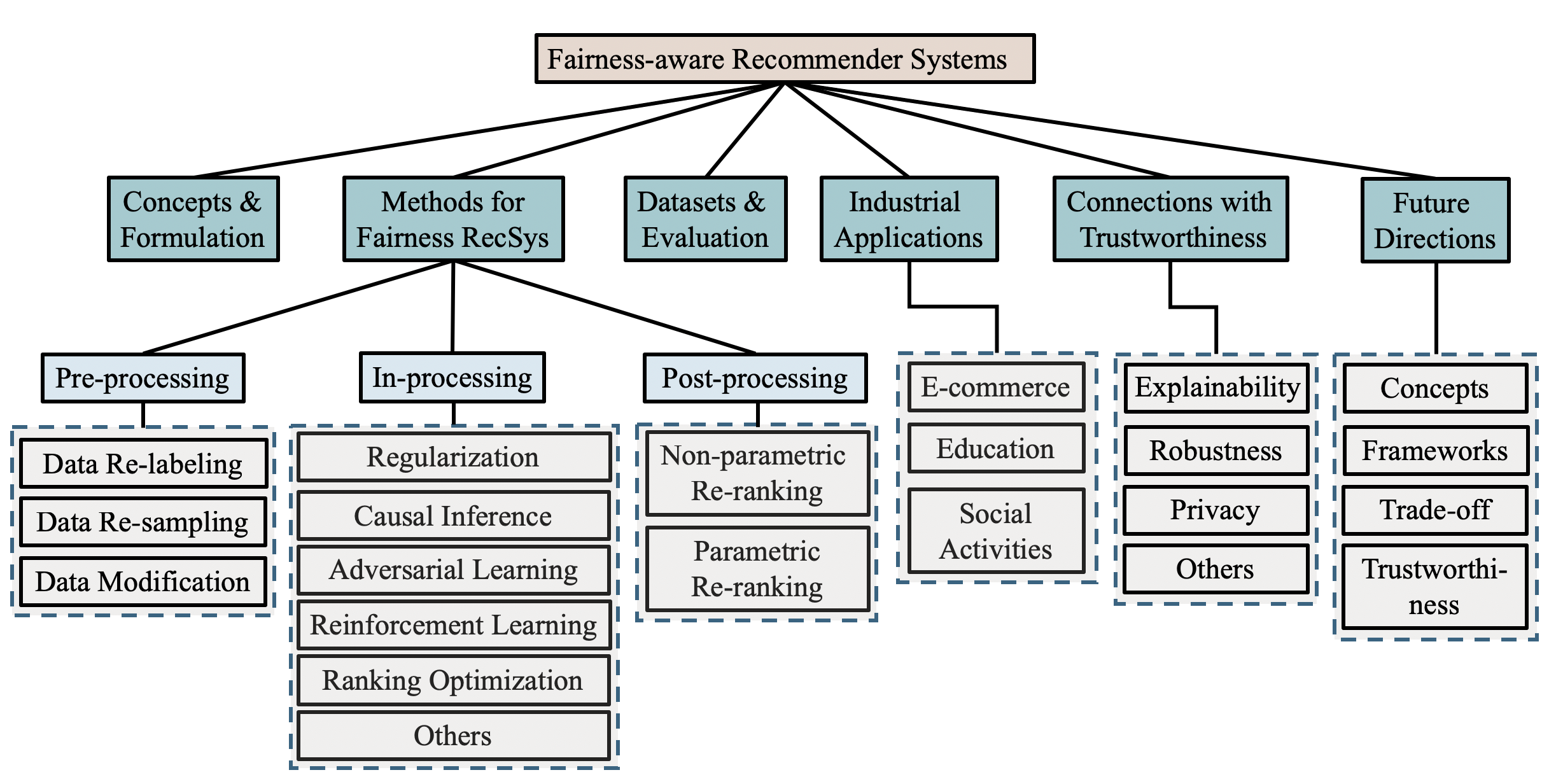} 
	\caption{The organizational layout of this survey.}
	\label{framework}
\end{figure*}

In this survey, we introduce the theory and practice of fairness in recommender systems. As a first step, we describe the concepts and formulations of recommender systems and fairness in Section \ref{conceptsss}. Then, we categorize and illustrate methods promoting fairness in different stages of recommender systems in Section \ref{methodsss} after carefully analyzing related literature regarding fairness in recommender systems. Specifically, these processing stages include pre-processing, in-processing, and post-processing. In Section \ref{dataset}, we collect and collate datasets and evaluation metrics used in literature exploring fairness in recommender systems. Following that, we discuss the industrial applications of recommendation systems that consider fairness in e-commerce, education, and social activities in Section \ref{application}. Moreover, Section \ref{connection} explores the connection between fairness and other ethical principles of a trustworthy recommender system such as explainability, robustness, privacy, and so on. Section \ref{Future} provides a big picture of future directions of fairness-aware recommender systems from different dimensions, including concepts, frameworks, trade-off, and trustworthiness. Last but not least, Section \ref{Conclusion} summarises this survey's importance and influence within the context of trustworthy recommender systems. Fig.\ref{framework} illustrates the organizational layout of this survey, as well as the logical connections between each section. The contributions of this survey can be enumerated as follows:

\begin{itemize}[leftmargin=0.4cm]
    \item \noindent\textbf{A Holistic Taxonomy of Fairness-aware RS Methods.} Compared to previous surveys, this study presents a more comprehensive taxonomy for methods that improve fairness in recommender systems. These methods are grouped in accordance with their roles in the three phases of implementing recommender systems: pre-processing, in-processing, and post-processing. 
    A useful and intuitive framework is provided for the adoption or understanding of these methods by interested researchers.
    
    \item \noindent\textbf{Building Connections between Fairness and other Ethical Principles.} This study demonstrates how fairness relates to other ethical principles in trustworthy recommendation systems. It encourages people to consider holistically while advocating fairness. A question such as, does promoting fairness affect other ethical dimensions of trustworthy recommendation systems, e.g., explainability and robustness, should be asked.
    
    \item \noindent\textbf{Evaluation of Existing Challenges and Future Direction.}
We highlight the existing limitations and challenges of existing fairness-promoting methods for recommender systems. In future works, these problems should be further considered and addressed. In particular, the definition of fairness is not consistent between studies, which can easily cause confusion. Additionally, despite improvements in fairness, existing methods neglect the relationship between fairness and other ethical dimensions of trustworthy systems, which can deteriorate other ethical metrics. 
\end{itemize}

\section{Concepts and Formulations}

\label{conceptsss}
\subsection{Recommender Systems} A recommender system serves as an information filtering system that learns and predicts the user's interest in an item \cite{DBLP:journals/kais/FreireC21}. According to relevant information such as items, users, and the interaction between them, the recommender system provides users with personalized services and recommends suitable items to them \cite{DBLP:journals/dss/LuWMWZ15}. Given a user $u\in U$, an item $v \in V$, and a recommender system $f(\cdot)$, where $U$ is a user set and $V$ is an item set, the goal of the recommender system is to learn an information filter $f(\cdot)$ to capture user preference.
The predicted score $y_{u,v}$ for the user's preferences in items is \cite{wu2020graph}:
\begin{equation}
\centering
y_{u,v} = {f}(u, v).
\end{equation}

\begin{figure*}[htpb]
	\centering
	\includegraphics[width=0.7\textwidth]{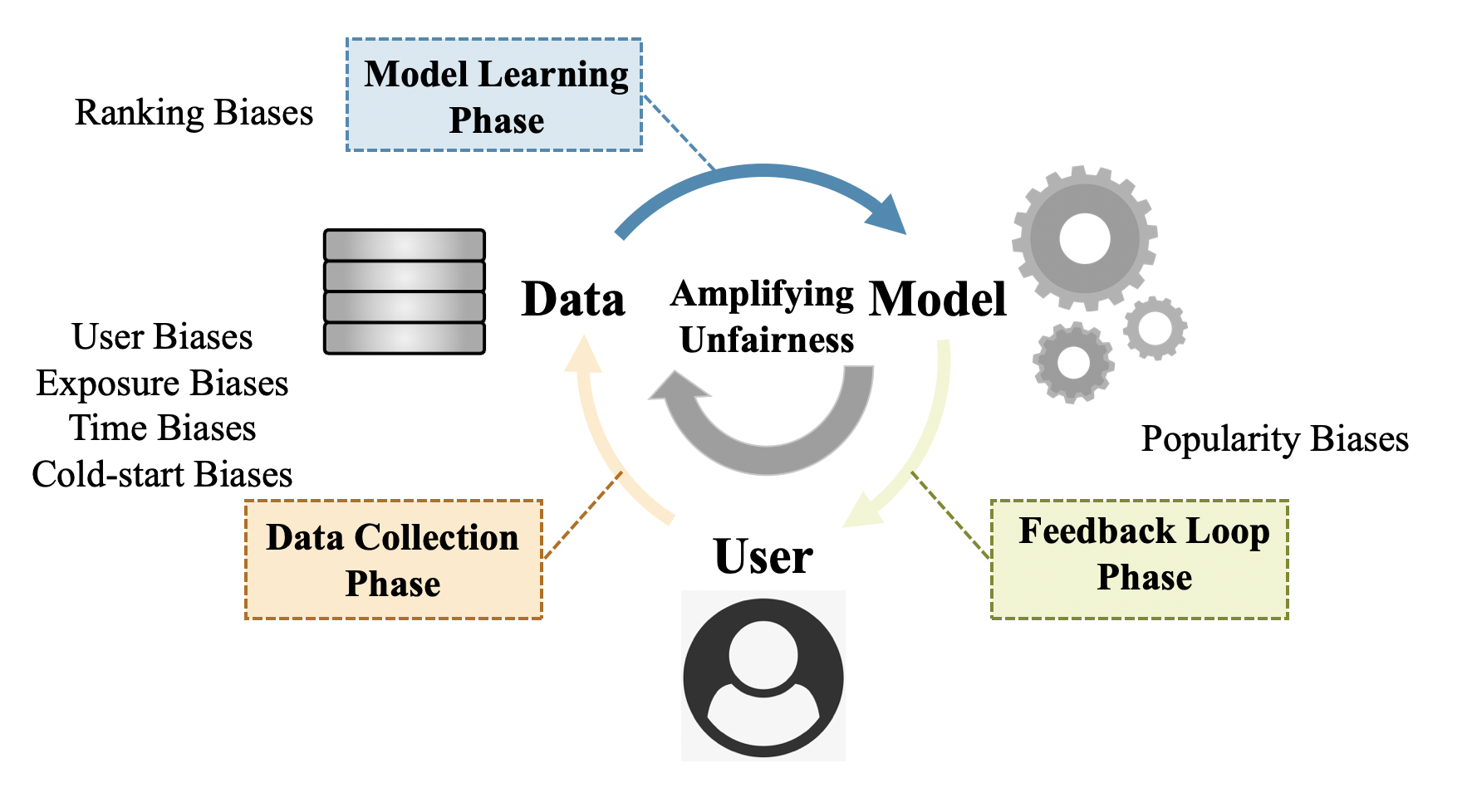} 
	\caption{The lifecycle of recommender systems. In the lifecycle, the data collection phase collects data from users; the model learning phase feeds data into the model for training; the loop phase provides recommendation results to users.
	In the lifecycle, biases in each phase will have an unfair impact on the recommender system. With the operation and feedback loop of recommender systems, existing biases can be potentially amplified in the following phases of recommender systems.}
	\label{bias}
\end{figure*}

The lifecycle of the recommender system can be abstracted into a feedback loop composed of users, data, and models. The loop consists of three phases as shown in Fig. \ref{bias}, including, a data collection phase, a model learning phase, and a feedback phase. The data collection phase is established between users and data, and it collects users' and items' attribute information, as well as user-item interaction data. The learning phase is built with the data and the model. In specific, this refers to the development of a recommendation model based on the collected data. 
A recommender system utilizes historical data to forecast the probability of an item being recommended to a user. This step delivers the suggested results to users to satisfy their information needs. This phase will impact users' future behavior and decisions. 

The recommender system can be divided into several types according to the different scenarios:

\begin{itemize}
    \item  \noindent\textbf{Session-based RSs} take each session as the input unit, capturing the user's short-term preferences and dynamic interests as reflected in session transitions \cite{DBLP:conf/www/JinWZSJLLP23}.
    \item  \noindent \textbf{Conversational-based RSs} are recommender systems that can interact with users in multiple rounds in real-time, eliciting users' dynamic preferences and taking actions based on their current needs \cite{DBLP:journals/aiopen/GaoLHRC21}.
    \item  \noindent \textbf{Content-enriched RSs} incorporate some auxiliary data related to users and items to enhance representation learning and semantic relevance. These auxiliary data may include textual content, knowledge graphs, etc \cite{wu2022survey}.
    \item  \noindent \textbf{Social RSs} are defined as any recommender system for the social media domain. This type of recommender systems improves recommendation performance by incorporating social relations into the recommender system \cite{Mooc}.
\end{itemize}
There are unique unfairnesses in different types of recommender systems, and we introduce these unfairness issues and the biases that generate unfairness in the follow-up.

\subsection{Fairness} 
Fairness is a concept that originated in sociology, economics, and law \cite{lamertz2002social,konow1996positive,michelman2013property}. Its definition in the Oxford English Dictionary is ``imperfect and just treatment or behavior without favoritism or discrimination''. 
In the context of recommender systems, fairness requires that recommender systems treat all users and items equally.
For example, in loan approval based on recommender systems \cite{musto2021fairness}, the systems are fair if the approval of an applicant seeking a loan is not influenced by the user's attributes (e.g., gender, race). We can develop fair recommender systems when bias or unfairness is eliminated from systems \cite{10.1145/3547333}. In this section, we first categorize causes of unfairness and then present common fairness expressions that can measure the existence of bias.

\subsubsection{Causes of Unfairness}
As depicted in Fig. \ref{bias}, the unfairness of a recommender system appears in its whole lifecycle, including the data collection phase, the model learning phase, and the feedback loop phase. According to the position of bias in the three phase, we divide bias in the recommender system into three categories, namely the data bias, model bias, and feedback bias. Subsequently, we will elucidate these biases by showing how they result in unfairness in recommender systems.


\noindent
\textbf{Data Bias.}
Currently, recommender systems are always trained on large datasets, which usually contain user information like user behaviors (e.g. incorrect clicks), sensitive attributes (e.g. gender), and the interaction between users and items.
However, training datasets potentially include various biases. 
These biases can lead to unfair recommendations, which may cause undesirable or even disastrous consequences for human life and society. 
Next, we will introduce some common biases in datasets.

\textit{User Bias}. 
User attribute bias is a common user bias. 
In recommender systems, some sensitive attributes 
like age \cite{salloum2021implementation}, geographical location \cite{lian2020geography}, gender \cite{ferraro2021break}, and profession \cite{giabelli2021skills2job} are important sources of information for recommender systems to understand user preferences. 
However, sometimes users' attributes (e.g., age, gender) can cause biased recommendation results. For example, age may be an effective feature in a music recommender system as they can recommend music to users according to their ages \cite{wu2022selective}. In general, youngsters have a higher preference for hip-hop music than senior people. However, sometimes users of a certain age group may wish to jump out of the cocoon of age information and explore different kinds of music. 
For the gender attribute, some researchers \cite{DBLP:journals/corr/DattaTD14, DBLP:conf/www/ImanaKH21,do2022online} observe the promotion of high-paying jobs varies by gender. 
In MOOC, a recommender system attempt to guide course resources to students \cite{DBLP:conf/ksem/ZhaoMJZ21}, the geographic location and nationality of teachers also influence which courses are recommended to students \cite{Mooc}. 


In addition to user attribute bias, user selection bias is also an important part of user bias. User selection bias refers to the behavioral bias when the user selects items, which usually exists in user explicit feedback. The existence of biases in user feedback data can lead to inconsistencies between user preferences and behavior records \cite{zheng2022cbr}. 
For instance, music streaming media recommender systems provide playlists based on the music preference of users \cite{DBLP:conf/recsys/JawaheerSK10a}, which can be affected by the user feedback bias. Specifically, during the recommendation process, the model is updated online based on user feedback. In a music recommender system, explicit feedback (e.g., ratings of items \cite{DBLP:conf/aaai/LiangP021}) can be the track marked "favorite" by the user. However, users may have wrongly clicked like tags, which will lead to explicit feedback bias \cite{zhang2022counteracting}.

\textit{Exposure Bias.} 
Exposure Bias refers to the situation where users only have access to a portion of the available item set.
\cite{DBLP:conf/icml/GuptaWLW21}. 
The exposure bias is often present in implicit feedback data, which usually refers to whether the user interacts with the item, including some purchases, clicks, and other behaviors \cite{DBLP:conf/sigir/ChengYLGLYLYC21}. 
For example, e-commerce sites like Amazon allow users to provide feedback on recommended items beyond the user interface by searching and browsing various product pages. The user's browsing behaviors may contain incorrect clicks, and these mistakes can make the recommender system misjudge the user's preference, which results in an unfair recommendation \cite{qin2020attribute}. 
In a music recommender system, a handful of popular artists may garner the majority of traffic, thereby underexposing less mainstream artists. If this bias remains unaddressed, recommender systems could adversely affect the experience of diverse users and items on the platform due to continuous interaction with biased recommendations, and thus the training of models using biased interactions in subsequent timeframes \cite{DBLP:journals/tois/MansouryAPMB22}.
In music streaming media recommender systems, the explicit feedback can be the track marked "favorite" by the user, and implicit feedback can be the number of times a track has been played. However, users may pay less attention to the music app while doing other activities, such as exercising, reading late at night, or commuting, and the music is looped multiple times, which can lead to an invisible feedback bias \cite{zhang2022counteracting}. 

\textit{Time Bias.} Some time-related recommendations, such as news recommendations, session-based recommendations, and job recommendations, heavily rely on direct user-item interactions to understand user preferences and provide specialized recommendations. However, freshly released data could unfairly overrepresent users' long-term interests, which is unfair to learn users' preferences. If we continue to disclose items at time $t+n$ according to the fairness limitations at time $t$, even though an item that is popular at time $t$ may no longer be popular at time $t+n$, we will neglect the long-term fairness dynamics \cite{ge2021towards}. This will ignore the long-term dynamic process of fairness, which leads to recommended bias. For example, the goal of job recommendations is to recommend job advertisements to job seekers. Job seekers prefer to click after seeing a new job advertisement. New job advertisements obtain higher click rates as a result of this behavior than older job advertisements. 
Consequently, jobs of longer-lasting professionals are less likely to be recommended. However, job seekers are driven by their long-term career aspirations when contemplating jobs. Advertising recommendations that are consistent with these preferences are more advantageous to job seekers \cite{10.1145/3357384.3358131}. Therefore, it is unfair to recommend recent advertisements to users without considering users' long-term interests.

\textit{Cold-start Bias.}
The above biases only consider unfairness in the case of warm-start recommendation. The primary source of unfairness in this instance is data biases (e.g., clicks or pageviews). However, the recommender system will meet a cold start problem when lacks data. 
When a new user or item enters the system, the cold-start problem occurs at which point the recommender system fails to provide personalized recommendations because it lacks sufficient data on user behaviors or item attributes \cite{DBLP:conf/recsys/CohenAKSN17}. Solving a cold-start problem is usually using prior knowledge from warm-start recommendations to train cold-start recommender systems. The cold-start bias refers to the bias of this prior knowledge in warm-start recommender systems.
These biases will be brought into the cold-start recommender system when training, and cause the unfairness phenomena \cite{vartak2017meta}. The unfairness phenomena can be particularly problematic because the unfairness caused by cold-start recommendations can persist and accumulate throughout the lifespan of the item, making it increasingly difficult to mitigate unfairness \cite{DBLP:conf/sigir/ZhuKNFC21}.

\noindent\textbf{Model Bias.}
The model learning phase uses the collected data in the data collection phase to train the recommendation model, the core of which is to deduce user preferences from past interaction data to predict the possibility of users choosing unvisited targets. Models with design flaws (e.g., ranking bias) may further magnify biases in the input data.

\textit{Ranking Bias.} 
Some loss functions can further exacerbate unfairness during the training of recommender systems. These loss functions affect the predicted score of the item, which in turn affects the recommendation ranking list of recommender systems. For example, Wan et al. \cite{wan2022cross} demonstrate that point losses (e.g., MSE loss) and pair-wise losses (e.g., BPR loss) are sensitive to popular items. These loss functions give popular items higher scores than unpopular items, which amplifies exposure bias during the training of recommender systems. Zhu et al. \cite{DBLP:conf/sigir/ZhuWC20} also prove that the BPR loss lacks the fairness constraint of equal opportunity ranking for reducing bias.

\noindent\textbf{Feedback Bias.}
A feedback loop exists in every recommender system, which provides recommendation results of a model to users for selection. In the feedback loop, users and a recommender system interact and co-evolve. Users' preferences and behaviors are updated through the recommender system, and the recommender system uses the updated data for self-reinforcing. This feedback loop mechanism not only generates bias, but also exacerbates the bias over time, resulting in a gradual deterioration of the fairness ecosystem. 
In the following paragraph, we introduce the popularity bias, which is a typical feedback bias.

\textit{Popularity Bias.} A few popular items are frequently recommended, while most others are disregarded. Users consume these recommendations, and their responses are recorded and added to the system. 
Over time, the recommender system recommends popular items to users, continuously collects user feedback on popular items, and adds them to the training set, making the data distribution more unbalanced,
which will result in more and more recommendation results focusing on popular items \cite{DBLP:conf/recsys/AbdollahpouriMB19}.
The existence of popularity bias will also constantly change the user's preference representation, making it challenging for the recommender system to capture the user's true preference. For example, Naghiaei et al. \cite{DBLP:conf/bias/NaghiaeiRD22} investigate the impact of popularity issues on book recommender systems. Their work shows that recommender systems tend to recommend popular items frequently, and there is a strong correlation existing between the popularity of books and the frequency with which they are recommended. Most books are not exposed to users by the recommender system, while popular books are highlighted more frequently.


\subsubsection{Fairness Expressions in Recommender Systems}
\label{sec:con:fiar:exp}


A recommender system can be employed in multiple scenarios, and the unfairness concerns in each scenario are different. Next, we present the expressions of fairness in recommender systems from different perspectives.

\textbf{Individual Fairness vs. Group Fairness.} 
Given two similar users $u_i,u_j \in U$, individually fairness-aware recommender systems hope to give similar prediction scores for samples $u_i$ and $u_i$ \cite{DBLP:conf/sigir/BiegaGW18}. For example, in a healthcare recommendation system, two patients exhibiting similar pathologies should receive recommendations of equivalent quality \cite{DBLP:journals/aiedu/MarrasBRF22}. 
Methods for individual fairness solve the problem of statistical equality through pairwise comparisons between similar users. 
The formal definition of individual fairness can be expressed as $f(u_i,v) \approx f(u_j,v).$
where $f(\cdot)$ is a recommender system, $v$ is an item.

Group fairness necessitates that protected groups receive treatment akin to that of advantaged groups \cite{DBLP:conf/sdm/PedreschiRT09}. 
As a typical group fairness, demographic parity \cite{DBLP:conf/www/LiCFGZ21} requires that the probability that the protected group (e.g., female group) is predicted to be a positive sample (e.g., job offers) is equal to the probability that the advantaged group (e.g., male group) is predicted to be a positive sample.
Given a protected group $U_i$ and an advantaged group $U_j$, the demographic parity of the recommender system can be formalised as
$Pr(y_{U_i,v} = 1 |U_i) = Pr(y_{U_j,v} = 1|U_j)$.

\textbf{Static Fairness vs. Dynamic Fairness.} 
Static fairness is defined as providing a short-term fair recommender system regardless of changes in the recommendation environment. 
Zhang et al. \cite{DBLP:journals/corr/abs-2109-03150} propose that the concept of recommendation fairness proposed by most methods is static fairness, i.e., the protected group is fixed during the recommendation process. As an example, traditional matrix-based recommendation strategies are aimed at maximization of users' immediate gratification, assuming that their preferences remain static. However, they potentially ignore users' long-term interests.


Dynamic fairness \cite{ge2021towards} is defined as considering dynamic factors in the environment to maintain fairness. Dynamic fairness is related to time bias. Some works argue that recommending an item that was recommended a long time ago should have the same probability as recommending an item that was recommended recently. Assuming $v_t$ represents an item appears in time $t$ and $v_{t+1}$ represents an item appears in in time $t+1$, the dynamic fairness can be defined as: $Pr(y_{u,v_t} = 1 | v_t) = Pr_t(y_{u,v_{t+1}} = 1 | v_{t+1})$. In real-world recommender systems, it can also be expressed as  $\frac{Pr(y_{u,v_t} = 1 | v_t)}{Pr(y_{u,v_{t+1}} = 1 |v_{t+1})} <\xi$, where $\xi$ is a slack factor used to adjust the dynamic fairness granularity for recommender systems.

\textbf{Single-sided Fairness vs. Multi-sided Fairness.} 
There are multiple stakeholders in a recommender system. 
Single-sided fairness refers to maintaining the interests (i.e., fairness) of a single role. For example, when recommending items to users, it only pays attention to whether the user has received a fair recommendation.
Multi-sided fairness refers to maintaining the interests of multiple roles \cite{naghiaei2022cpfair}. For example, consumers expect the recommender system to fairly recommend products they are interested in, while suppliers wish their products to be fairly exposed to consumers. 
Multi-sided fairness aims to maintain the fairness of both consumers and suppliers.
A typical multi-sided fairness is to increase diversity \cite{DBLP:conf/cikm/MehrotraMBL018} in recommendation results, ensuring that the average exposure of items across multiple aspects is balanced \cite{DBLP:conf/kdd/SinghJ18}. In addition, it's plausible to enforce proportional exposure of both groups relative to their average utility, i.e.,
$\frac{\sum_{i=1}^{|V_i|}Pr(y_{u,V_i} = 1 | V_i)}{|V_i|} = \frac{\sum_{j=1}^{|V_j|}Pr(y_{u,V_j} = 1 | V_j)}{|V_j|},$
where $V_i$ and $V_j$ are two different item groups.

\textbf{Others.} 
Other common perspectives of fairness in recommender systems are: 
\begin{itemize}
    \item \textit{Ranking Fairness.} The fairness of sorting is usually reflected in statistical equality or equality of opportunity. A simple interpretation of statistical parity in ranking is the assurance that the proportion of protected individuals appearing within a ranking prefix exceeds a predetermined threshold \cite{DBLP:conf/ssdbm/YangS17}. 
    In general, ranking fairness asks that similar items or groups of items receive similar visibility, they appear at similar positions in the ranking \cite{Linkin}. 
    \item \textit{Causal Fairness.} Most recommender systems are based on statistics; however, this ignores causality in the original data. Unlike individual fairness and equal opportunity, causal fairness not only considers observational data but also incorporates additional causal relationships within them \cite{zhang2021causal}. 
    Counterfactual fairness is a typical causal fairness. It demands that the predictions of recommender systems remain consistent in both the factual and the counterfactual world \cite{DBLP:conf/sigir/Tavakol20,DBLP:journals/csur/MehrabiMSLG21}.
    
    \item \textit{Long-term Fairness.} Long-term fairness is affected on time \cite{ge2022toward} and considers the long-term impact of fairness interventions \cite{DBLP:conf/aies/AkpinarDN022}. It is similar to dynamic fairness in that it considers the long-term impact on user preferences \cite{DBLP:conf/icml/MladenovCBSZB20}. 
    \item \textit{Envy-freeness Fairness.} Do et al. \cite{DBLP:conf/aaai/DoCAU22} propose a criterion of envy-freeness fairness, stating that each user should prefer their own recommendations over those of other users.
\end{itemize}
Other fairness concepts, like Rawlsian Maximin Fairness or Maximin-Shared Fairness, can be found in recent literature on fairness \cite{10.1145/3547333}.



\section{Methods for Fairness-aware Recommender Systems}

\label{methodsss}
\begin{figure*}[ht!]
	\centering
	\includegraphics[width=0.7\textwidth]{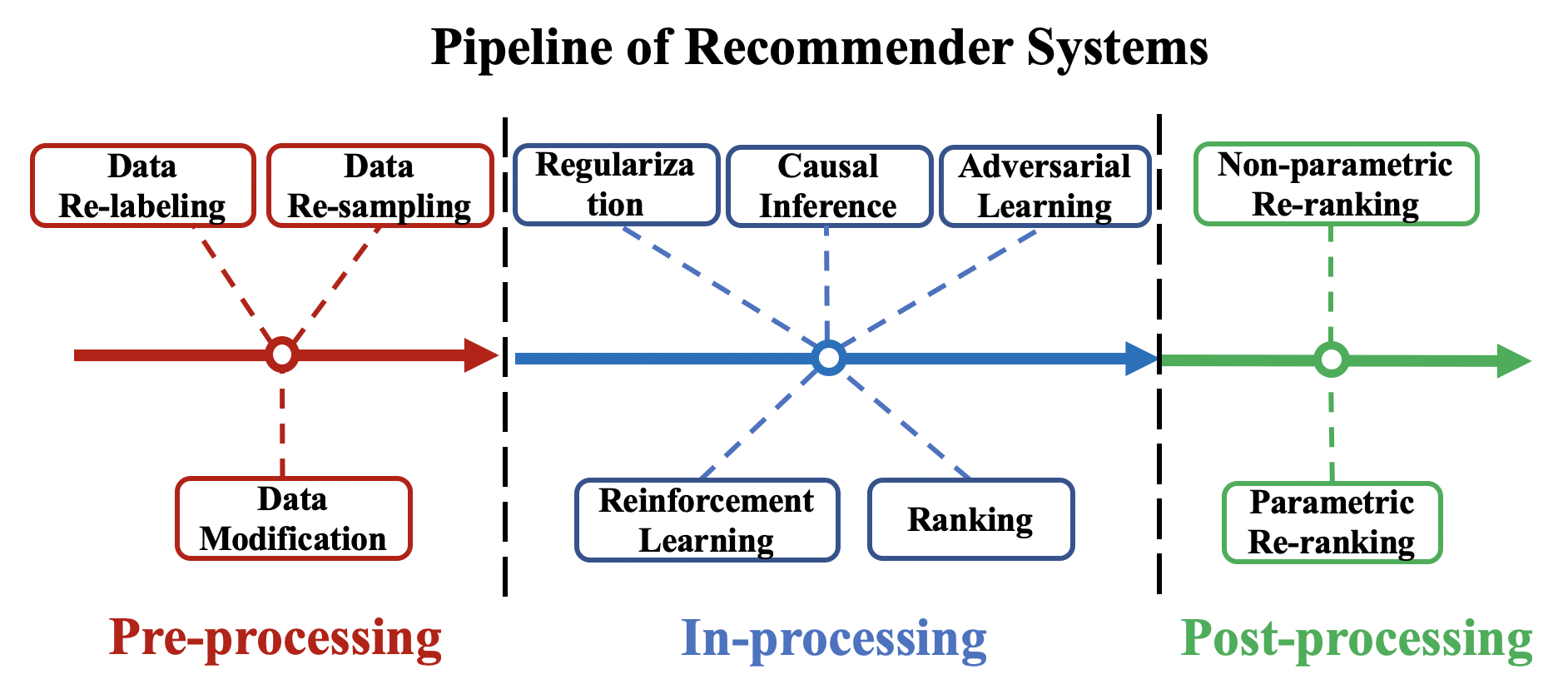} 
	\caption{Fairness-aware methods in the three stages. The pre-processing stage encompasses several methods like data re-labeling, data re-sampling, and data modification. These pre-processing methods are independent of recommendation models. The in-processing stage is mainly to improve the fairness of the model from a method perspective. The in-processing stage involves various techniques, including regularization-based methods, causal-inference-based methods, adversarial-learning-based methods, reinforcement-learning-based methods, and ranking-optimization-based methods. The post-processing stage also employs model-independent methods, treating the model as a black box and solely processing the model's output results. The post-processing methods include non-parametric re-ranking methods and parametric re-ranking methods.}
	\label{3stages}
\end{figure*}

In this section, we summarize methods for improving the fairness of recommender systems and present them in three stages of implementation, including pre-processing, in-processing, and post-processing (as shown in Fig. \ref{3stages}). In addition, we summarize the fairness-aware recommendation methods in Table \ref{methodtable}, specifically including the biases the methods can address, the types of fairness-aware recommendation methods, and other important information.

\subsection{Pre-processing Methods for Fairness Recommender Systems}

In the pre-processing stage, fairness-enhancing methods are used to minimize biases in the training datasets. It is possible for these biases to be amplified throughout the lifecycle of a recommender system, resulting in unfair recommendations for users. We divide these fairness-aware pre-processing methods into three categories according to data debiasing methods.

\subsubsection{Data Re-labeling} 
Re-labeling changes the labels of the training dataset to remove biases in the input data. Some recommender systems predict user preferences using binary implicit feedback labels, like ``click" or ``not click". However, there may be noise in the feedback labels, i.e., clicks do not necessarily represent positive feedback, and missing clicks do not necessarily represent negative feedback. These ``noisy" labels can lead to a drop in model performance.
Wang et al. \cite{DBLP:conf/mm/WangX0CH21} design a self-supervised re-labeling framework for noise in implicit feedback. The framework dynamically generates pseudo labels for user preferences to mitigate noise in both observed and unobserved feedback data.
\begin{table*}
\centering
\setlength{\abovecaptionskip}{0pt}
\setlength{\belowcaptionskip}{0pt}
\small
\caption{Summary of Methods. We classify existing methods based on the phase of implementation, the technology utilized, the bias issues encountered, the suggested application scenarios, and the official method names.}
\label{methodtable}
\begin{center} 

\begin{threeparttable}

\begin{tabular}{@{}ccllp{7cm}@{}}
\toprule
\multicolumn{1}{l}{\textbf{Stage}}                                          & \multicolumn{1}{l}{\textbf{Technologies}}                                          & \textbf{Bias}                  & \textbf{Background} & \textbf{Methods}                                                                                                                                                                                                                                                                                                                                                                                                                                                                            \\ \midrule
\multirow{4}{*}{\begin{tabular}[c]{@{}c@{}}Pre-\\ processing\end{tabular}}  & \multirow{2}{*}{Data Re-labeling}                                                  & Popularity Bias                & General RSs         & IFNA*{\cite{DBLP:conf/mm/WangX0CH21}}                                                                                                                                                                                                                                                                                                                                                                                                                                                       \\ \cmidrule(l){3-5} 
                                                                            &                                                                                    & User Bias                      & General RSs         & DPTC*{\cite{DBLP:journals/kais/KamiranC11}}                                                                                                                                                                                                                                                                                                                                                                                                                                                 \\ \cmidrule(l){2-5} 
                                                                            & Data Re-sampling                                                                   & User Bias                      & General RSs         & MPML*{\cite{DBLP:journals/ci/EstabrooksJJ04}}, HPO{\cite{DBLP:conf/sac/MontanariBC22}}, RNS{\cite{DBLP:conf/ijcai/DingQ00J19}}, HFD*{\cite{celis2016fair}}                                                                                                                                                                                                                                                                                                                                  \\ \cmidrule(l){2-5} 
                                                                            & Data Modification                                                                  & User Bias                      & General RSs         & \begin{tabular}[c]{@{}l@{}}HURR*{\cite{DBLP:conf/sigir/SachdevaM20}}, CFAI*{\cite{DBLP:journals/tosem/WangYWHW22}}, RSNMF{\cite{DBLP:journals/tii/LuoZXZ14}}, \\ EPTB*{\cite{ DBLP:journals/ijon/YuanLS18}}\end{tabular}                                                                                                                                                                                                                                                                    \\ \midrule
\multirow{19}{*}{\begin{tabular}[c]{@{}c@{}}In-\\ processing\end{tabular}}  & \multirow{4}{*}{Regularization}                                                    & Cold-start Bias                & Cold-start RSs      & CLOVER{\cite{wei2022comprehensive}}                                                                                                                                                                                                                                                                                                                                                                                                                                                         \\ \cmidrule(l){3-5} 
                                                                            &                                                                                    & Exposure Bias                  & Social RSs          & MinDiff{\cite{DBLP:journals/corr/abs-1910-11779}}, FRRPC*{\cite{DBLP:conf/kdd/BeutelCDQWWHZHC19}}, SERec{\cite{DBLP:conf/aaai/WangZYZ18}}, IDLR*{\cite{DBLP:conf/flairs/WasilewskiH16}}                                                                                                                                                                                                                                                                                                     \\ \cmidrule(l){3-5} 
                                                                            &                                                                                    & Popularity Bias                & General RSs         & FARL*{\cite{kiswanto2018fairness}}, CPBLR*{\cite{DBLP:conf/recsys/AbdollahpouriBM17}}                                                                                                                                                                                                                                                                                                                                                                                                       \\ \cmidrule(l){3-5} 
                                                                            &                                                                                    & User Bias                      & General RSs         & \begin{tabular}[c]{@{}l@{}}CFIFD*{\cite{DBLP:conf/icdm/HuKV08}}, FFFU*{\cite{DBLP:conf/wsdm/RastegarpanahGC19}}, SLIM{\cite{DBLP:conf/fat/BurkeSO18}}, FaiRecSys{\cite{DBLP:journals/ijdsa/EdizelBHPT20}}, \\ MMIUR*{\cite{DBLP:conf/sigir/ZhuWC20}}, F2VAE{\cite{DBLP:conf/sac/BorgesS22}}, FairRec{\cite{DBLP:conf/aaai/WuWWH021}}, \\ IURPF*{\cite{DBLP:journals/umuai/BorattoFM21}}, FRRP*{\cite{DBLP:conf/kdd/BeutelCDQWWHZHC19}},  FATBR*{\cite{DBLP:conf/cikm/ZhuHC18}}\end{tabular} \\ \cmidrule(l){2-5} 
                                                                            & \multirow{6}{*}{\begin{tabular}[c]{@{}c@{}}Causal\\ Inference\end{tabular}}        & Cold-start Bias                & Cold-start RSs      & CLOVER{\cite{wei2022comprehensive}}                                                                                                                                                                                                                                                                                                                                                                                                                                                         \\ \cmidrule(l){3-5} 
                                                                            &                                                                                    & Exposure Bias                  & Social RSs          & DHRS*{\cite{DBLP:conf/www/SunKNS19}}, ABPUL*{\cite {10.1145/3394486.3403285}}                                                                                                                                                                                                                                                                                                                                                                                                                \\ \cmidrule(l){3-5} 
                                                                            &                                                                                    & Popularity Bias                & General RSs         & IANP*{\cite{DBLP:conf/sigir/0002WLCZDWSLW22}}, PDA{\cite{zhang2021causal}},  MACR*{\cite{wei2021model}}, DANCER{\cite{huang2022different}}                                                                                                                                                                                                                                                                                                                                                  \\ \cmidrule(l){3-5} 
                                                                            &                                                                                    & \multirow{3}{*}{User Bias}     & Content RSs         & FairTED{\cite{acharyya2020fairyted}}                                                                                                                                                                                                                                                                                                                                                                                                                                                        \\ \cmidrule(l){4-5} 
                                                                            &                                                                                    &                                & General RSs         & \begin{tabular}[c]{@{}l@{}}CSBRS*{\cite {DBLP:conf/wsdm/0003ZS021}}, SHT{\cite{DBLP:conf/kdd/Xia0Z22}},  RTDLE*{\cite{schnabel2016recommendations}}, TPFB*{\cite{10.1145/3404835.3462966}}, \\ AdaRequest{\cite{DBLP:conf/kdd/QianXLZJLZC022}}, InvPref{\cite {wang2022invariant}}, DeSCoVeR{\cite{rajanala2022descover}}\end{tabular}                                                                                                                                                      \\ \cmidrule(l){4-5} 
                                                                            &                                                                                    &                                & Session-based RSs   & AIPB*\cite{lesota2021analyzing}                                                                                                                                                                                                                                                                                                                                                                                                                                                            \\ \cmidrule(l){2-5} 
                                                                            & Adversarial L                                                                      & User Bias                      & General RSs         & FairRec{\cite{DBLP:conf/aaai/WuWWH021}}, FRFC{\cite{DBLP:journals/kbs/LiuZZL000F22}}, CGWG*{\cite{DBLP:journals/corr/abs-2209-09592}}                                                                                                                                                                                                                                                                                                                                                       \\ \cmidrule(l){2-5} 
                                                                            & \multirow{2}{*}{\begin{tabular}[c]{@{}c@{}}Reinforcement \\ Learning\end{tabular}} & Exposure Bias                  & Social RSs          & FCPO{\cite{ge2021towards}}, MORL{\cite{ge2022toward}}                                                                                                                                                                                                                                                                                                                                                                                                                                       \\ \cmidrule(l){3-5} 
                                                                            &                                                                                    & Popularity Bias                & Conversational RSs  & Popcorn{\cite{DBLP:conf/cikm/FuXGMZ21}}                                                                                                                                                                                                                                                                                                                                                                                                                                                     \\ \cmidrule(l){2-5} 
                                                                            & \multirow{3}{*}{Ranking}                                                           & Exposure Bias                  & Social RSs          & RDC{\cite {liu2022rating}}                                                                                                                                                                                                                                                                                                                                                                                                                                                                  \\ \cmidrule(l){3-5} 
                                                                            &                                                                                    & Popularity Bias                & General RSs         & CPR{\cite{wan2022cross}}                                                                                                                                                                                                                                                                                                                                                                                                                                                                    \\ \cmidrule(l){3-5} 
                                                                            &                                                                                    & User Bias                      & General RSs         & MMIUR*{\cite{DBLP:conf/sigir/ZhuWC20}}                                                                                                                                                                                                                                                                                                                                                                                                                                                      \\ \cmidrule(l){2-5} 
                                                                            & \multirow{3}{*}{Others}                                                            & Exposure Bias                  & Session-based RSs   & CLRec{\cite {zhou2021contrastive}}, SAR-Net{\cite{DBLP:journals/corr/abs-2110-06475}}                                                                                                                                                                                                                                                                                                                                                                                                       \\ \cmidrule(l){3-5} 
                                                                            &                                                                                    & User Bias                      & Conversational RSs  & CUAB*{\cite{zhang2022counteracting}}                                                                                                                                                                                                                                                                                                                                                                                                                                                        \\ \cmidrule(l){3-5} 
                                                                            &                                                                                    & User Bias                      & Social RSs          & FairSR{\cite{li2022fairsr}}                                                                                                                                                                                                                                                                                                                                                                                                                                                                 \\ \midrule
\multirow{3}{*}{\begin{tabular}[c]{@{}c@{}}Post-\\ processing\end{tabular}} & \multirow{2}{*}{Non-parametric}                                                    & \multirow{2}{*}{Exposure Bias} & Cold-start RSs      & GEN{\cite{DBLP:conf/sigir/ZhuKNFC21}}                                                                                                                                                                                                                                                                                                                                                                                                                                                       \\ \cmidrule(l){4-5} 
                                                                            &                                                                                    &                                & General RSs         & CPFair{\cite{naghiaei2022cpfair}}, FairMatch{\cite{mansoury2020fairmatch}}, TFROM{\cite{DBLP:journals/corr/abs-2104-09024}}                                                                                                                                                                                                                                                                                                                                                                 \\ \cmidrule(l){2-5} 
                                                                            & Parametric                                                                         & Exposure Bias                  & General RSs         & HyperFair{\cite{DBLP:journals/corr/abs-2009-08952}}                                                                                                                                                                                                                                                                                                                                                                                                                                         \\ \bottomrule
\end{tabular}

\end{threeparttable}

\end{center}
\begin{threeparttable}
\begin{tablenotes}
    \footnotesize \item[] In this table, ``*'' indicates the method has no specific name. We named it with the abbreviation of its article name. 
\end{tablenotes}
\end{threeparttable}
\vspace{-0.5cm}
\end{table*}


\subsubsection{Data Re-sampling} 
Unbalanced data distribution may lead to a decrease in the training effect of recommender systems.
Montanari et al. \cite{DBLP:conf/sac/MontanariBC22} design a data sampling algorithm to ensure that the sampling is uniform and not affected by the distribution. They randomly select a certain percentage of users and delete their interaction information. This type of re-sampling method reduces the size of the dataset for approximately the same proportion of users and maintains the dynamic nature of user profiles.
Celis et al. \cite{celis2016fair} propose a subsampling method that proportionally subsamples the dataset based on different sensitive attributes.
Ding et al. \cite{DBLP:conf/ijcai/DingQ00J19} design a negative sampler that creates data resembling
generates data similar to the exposure data through feature-matching techniques instead of selecting directly from exposure data. 
The sampler forces the distribution of positive and negative data to be balanced by adding negative samples.

\subsubsection{Data Modification}
The main idea of data modification is to augment or modify the biased data to reduce the bias. For example, some recommender systems may utilise textual data (e.g. job recommender systems, news recommender systems), which may be missing or have errors. This incomplete or noisy data can lead to data bias. 
To solve the noisy data problems, Wang et al. \cite{DBLP:journals/tosem/WangYWHW22} implemented natural language processing (NLP) pre-processing techniques to modify the training data.
Specifically, in a crowdtesting recommendation (e.g., recommending software testing tasks to professionals), they first perform standard word segmentation for each document. Then, they remove stop words and apply synonym substitution to reduce noise. In addition, they construct a descriptive term list and perform term filtering for each document. 
Similarly, Sachdeva et al. \cite{DBLP:conf/sigir/SachdevaM20} use a similar NLP approach as Wang et al. In addition to handling noisy data, some works focus on missing data. Collaborative filtering-based recommenders typically model user preferences as a user-item rating matrix. Since it is not possible for the user to interact with all the items. The rating matrix is thus high-dimensional and sparse (HiDS), with many missing data representing the user's unobserved preferences.
Some works \cite{DBLP:journals/ijon/YuanLS18, DBLP:journals/tii/LuoZXZ14} apply a latent factor-based model, which provides a good job of handling the high-dimensional and sparse (HiDS) matrices, to process missing data.

\begin{figure*}[htpb]
	\centering
	\includegraphics[width=0.8\textwidth]{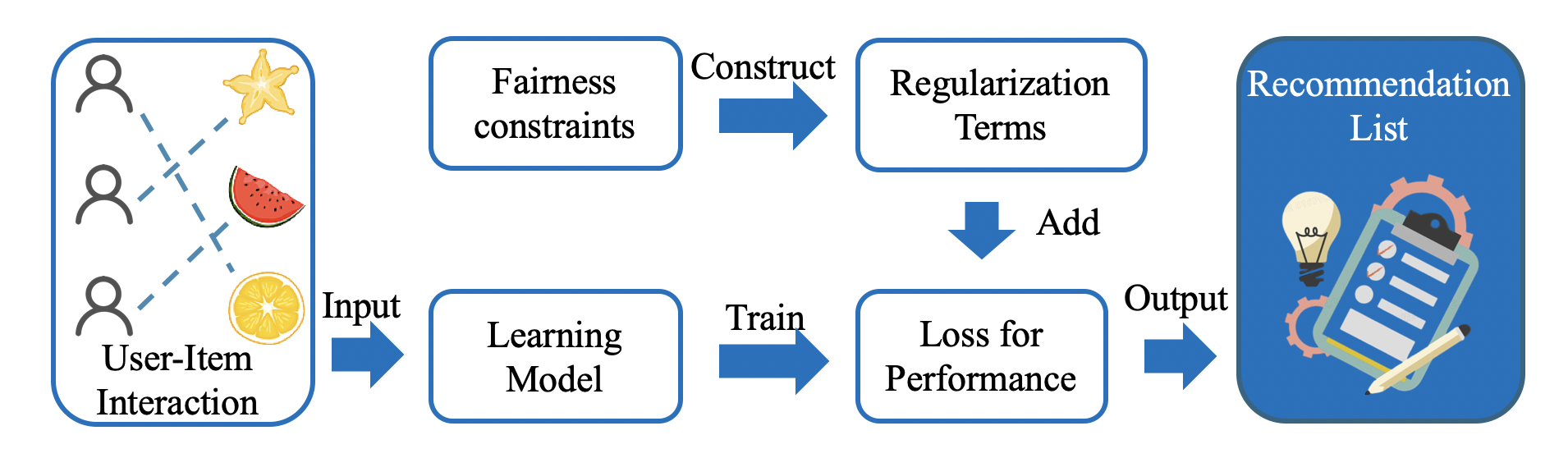} 
	\caption{Regularization-based fairness-aware methods. }
	\label{regularization}
\end{figure*}

\subsection{In-processing Methods for Fair Recommender Systems}
Currently, most recommender systems (e.g., collaborative filtering methods \cite{DBLP:conf/sigir/0002WLCZDWSLW22}) are devised to extract user preferences by learning the correlation in the training data \cite{DBLP:conf/kdd/Xia0Z22}. 
However, people potentially suffer from unfairness services when using these correlation-oriented recommender systems, such as Simpson's paradox \cite{simpson1951interpretation}, popularity bias \cite{zhang2021causal}, user-oriented bias \cite{DBLP:conf/kdd/QianXLZJLZC022}, cold-start bias \cite{wei2022comprehensive}, to name only a few.
In addition to removing bias via pre-processing methods, many works design fairness-aware methods to alleviate or even eliminate unfairness during model training of recommender systems. In-processing fairness-aware methods aim to learn bias-free models. 
In this section, we classify these works into five categories including regularization-based methods, casual-inference-based methods, adversarial-learning based, reinforcement-learning-based methods, ranking methods, and others.

\subsubsection{Regularization and Consternation for Fairness}
To alleviate unfairness in recommender systems, regularization penalizes the predicted recommendation score in accordance with the fairness evaluation \cite{DBLP:journals/corr/abs-2010-03240}. Regularization terms are widely used in various recommendation scenarios to reduce bias, such as user-oriented bias \cite{wei2022comprehensive}, group bias \cite{DBLP:conf/kdd/BeutelCDQWWHZHC19}, exposure bias \cite{DBLP:conf/aaai/WangZYZ18}, popularity bias \cite{kiswanto2018fairness}, etc.

In general, a regularization term is usually regarded as an additional loss focusing on promoting fairness. The term is added to the main loss, which is mainly responsible for improving the recommendation performance. Specifically, the total loss used for recommendation training is the sum of the performance loss and the regularization term:
\begin{equation}
    \mathcal{L} = \mathcal{L}_{per}(\mathbf{u},\mathbf{v}) + \mathcal{L}_{reg}(\mathbf{\theta}),
\end{equation}
where $L$ is the total loss for training, $L_{per}(\cdot)$ is the loss for optimizing the recommendation performance, $L_{reg}(\cdot)$ is a customized loss function, and $\theta$ refers to all possible parameters related to fairness evaluation.
Fig. \ref{regularization} shows a general framework for regularization-based fairness-aware methods.
According to the composition of the regularization term, we can divide the regularization into three categories, including norm-based regularization terms, matrix-based regularization terms, and pair-wise regularization terms.

\begin{figure*}[htpb]
	\centering
	\includegraphics[width=0.8\textwidth]{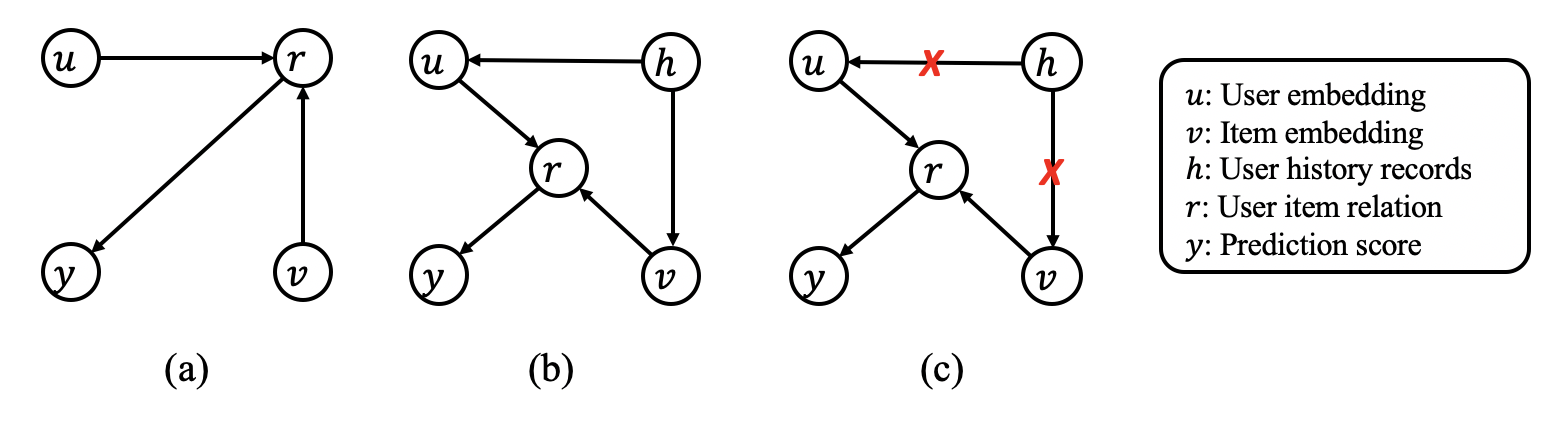} 
	\caption{
	A causal graph is a directed acyclic graph, wherein each node stands for a random variable, and the directed edges indicate causal relationships. In a recommender system, the recommendation process can be briefly described as calculating the matching score $r$ between user $u$ and item $v$, to predict the recommendation score $Y$. 
	As shown in (a), where $u$ is the user's representation, $v$ is the item's representation, $h$ represents the user's history, and $y$ is the prediction score. 
	$u\rightarrow r$ denotes that $u$ is an inducement of $r$, and $R$ has been effected by a direct causal from $u$. Similarly, there is no arrow between $u$ and $y$, indicating that $u$ has no direct causal relationship with $y$. 
	The popularity bias, a prevalent form of unfairness in existing recommender systems, results from users repeatedly clicking recommended items and then recommender systems always advocating for these items. As shown in (b), the history $h$ influences $u$ and $v$, which determine $r$. (c) shows a method to decrease the effect of $h$ on $u$ and $v$.}
	\label{causalgraph}
\end{figure*} 

\textbf{Norm-based Regularization.} 
The norm calculates the distance between raw features and generated embeddings. It is used to evaluate the deviation of the model of learnable recommender systems. 
Burke et al. \cite{DBLP:conf/fat/BurkeSO18} propose the possibility of using $l_1$ norm, and $l_2$ norm as regularization terms. Among multi-sided unfairness issues, user neighborhoods can constrain deviance in their opinion. 
The regularization term can be described as:
\begin{equation}
    \mathcal{L}_{reg} = \lambda_1\parallel W\parallel ^1 + \frac{\lambda_2}{2}\parallel W\parallel ^2 + \frac{\lambda_3}{2} \sum_{i}^n (b_i)^2,
\end{equation}
where $W$ is a user-user weight matrix, $\parallel \cdot \parallel ^1$ is $l_1$ norm, $\parallel \cdot \parallel ^2$ is $l_2$ norm, and $b_i$ is a neighborhood balance regularization for reducing the probability of user neighborhoods forming. 
$b_i$ is the squared difference between the weights of the protected users versus the unprotected users. Protected users are usually including sensitive attributes, and unprotected users do not have sensitive attributes.
The works \cite{DBLP:conf/wsdm/RastegarpanahGC19, DBLP:conf/cikm/ZhuHC18} employ similar techniques to alleviate group unfairness.
Hu et al. \cite{DBLP:conf/icdm/HuKV08} use the $l_2$ norm of user embeddings and item embeddings as a regularization term.

\textbf{Matrix-based Regularization.} 
Some regularization terms are in the form of matrices.
Abdollahpouri et al. \cite{DBLP:conf/recsys/AbdollahpouriBM17} introduce a matrix-based regularization LapDQ to reduce the popularity bias. The LapDQ regularizer is defined as :
\begin{equation}
    \mathcal{L}_{reg} = tr(V^TL_DV),
\end{equation}
where $V$ is the item embedding matrix, $tr(\cdot)$ is the trace function, and $L_D$ is the Laplacian of the dissimilarity matrix $D$. 
Wasilewski et al. \cite{DBLP:conf/flairs/WasilewskiH16} and Edizel et al. \cite{DBLP:journals/ijdsa/EdizelBHPT20} also use this type of regularizer for ranking unfairness and user bias. 

\textbf{Correlation-based Regularization.}
This type of method mainly exploits correlation to reduce bias.
Beutel et al. \cite{DBLP:conf/kdd/BeutelCDQWWHZHC19} propose a correlation-based pair-wise regularization term to balance the clicked and unclicked item, which is defined as:
\begin{equation}
    \mathcal{L}_{reg} = |Corr(A,B)|,
\end{equation}
where $Corr(\cdot)$ calculates the absolute correlation of two random variables, $A$ and $B$ are two random variables. 
The specific meaning of $A$ in this work is the residuals between clicked and unclicked items. Variable $B$ means the correlation between group users of the clicked items. The model is penalized if it predicts that one group clicked on an item more than the other group did. 
Prost et al. \cite{DBLP:journals/corr/abs-1910-11779} propose a MinDiff formulation based on the above method. MinDiff minimizes the correlation of predicted probability distribution and the distribution between the clicked items and unclicked items. 

\textbf{Others.}
To further eliminate data bias in the model training, Wu et al. \cite{DBLP:conf/aaai/WuWWH021} propose an orthogonality regularization to orthogonalize the unbiased user embeddings to the biased user embeddings. It thus distinguishes between embeddings that are unbiased and those that are biased.
For each user $u$, the orthogonality regularization can be defined as:
\begin{equation}
    \mathcal{L}_{reg}(u^b,u^d)=|\frac{\mathbf{u}^b\cdot\mathbf{u}^d}{\parallel \mathbf{u}^b \parallel \cdot \parallel\mathbf{u}^d\parallel}|,
\end{equation}
where $\mathbf{u}^b$ and $\mathbf{u}^d$ are the bias-aware and bias-free embeddings, respectively. For group fairness, Boratto et al. \cite{DBLP:journals/umuai/BorattoFM21} customized a special regularization term to reduce the sensitive attribute in group fairness.
\begin{equation}
    \mathcal{L}_{reg} = (\frac{\sum_i^n f(u,v)\cdot \mathcal{S}(v)}{\sum_i^n f(u,v)} - C)^2,
\end{equation}
where $\mathcal{S}(v)$ represents the percentage of users who have interacted with item $v$, 
$C$ represents the proportion of interactions between groups with sensitive attributes and a certain type of item in all interactions.
This regularized optimization implies that the model is penalized if the difference between correlation and contribution of the population is significant. 
Zhu et al. \cite{DBLP:conf/sigir/ZhuWC20} use Kullback-Leibler Divergence to normalize user prediction scores to a normal distribution, reducing the ranking unfairness in model training.

\subsubsection{Causal Inference for Fairness} 
\label{sec:methods:in:ci}
The causal inference in artificial intelligence explores the causal relationships between variables, i.e., how one variable determines another variable.
In this survey, \textit{causal inference for fairness} represent methods (e.g. inverse propensity score \cite {10.1145/3394486.3403285, xiao2022towards}) that trace to the source of bias and then mitigate unfairness through causal inference \cite{10.1145/3404835.3462966, wang2022invariant, wei2021model}. As shown in Fig. \ref{causalgraph}, causal graphs, which visually represent causal relationships between variables in a recommender system, are used to analyze the causes of unfairness. In this section, based on the usage of causal graphs, we categorise current causal inference methods for fairness into inverse propensity scoring, backdoor adjustment, and counterfactual inference.

\textbf{Inverse Propensity Score.} 
By analyzing the causes of bias in the causal graph, the inverse propensity score (IPS) reweights samples in order to reduce the influences of biased samples, without changing the causal relationship between variables \cite{DBLP:conf/www/SunKNS19}. In Fig. \ref{causalgraph} (b), for instance, the confounding variable $h$ affects both $u$ and $i$, the probability distribution $\hat{Pr}(h,u,v,r)$ in this figure can be expressed as
\begin{equation}
    \hat{Pr}(h,u,v,r)=Pr(r|u,v,h)Pr(u|h)Pr(v|h),
\end{equation}
while in the absence of the confounding variable (i.e., Fig. \ref{causalgraph} (c)), the distribution is:
\begin{equation}
    Pr(h,u,v,r)=Pr(r|u,v,h)Pr(u)Pr(i)Pr(h).
\end{equation} 
To remove the bias brought by the confounding variable $h$, IPS methods expect $\hat{Pr}(h,u,v,r) = Pr(h,u,v,r)$, which can be achieved through multiplying $\hat{Pr}(h,u,v,r)$ by a propensity weight, i.e.,  
\begin{equation}
    Pr(h,u,v,r)= \frac{Pr(h)}{Pr(u|h)Pr(v|h)}\hat{Pr}(h,u,v,r).
\end{equation}

Recently, some methods using IPS have been studied to conduct debiasing. 
A propensity framework proposed by \cite {10.1145/3394486.3403285} makes propensity estimations for improving exposure fairness in implicit feedback scenarios. 
Xiao et al. \cite {xiao2022towards} fuse a deep variational information bottleneck approach with a propensity score to develop an unbiased learning algorithm. 
Since user preferences might drastically change over time, Huang et al. \cite{huang2022different} exploit a dynamic inverse propensity score for debiasing dynamic popularity biases. 
To address user behavior bias, Wang et al. \cite {DBLP:conf/wsdm/0003ZS021} use unbiased data to introduce propensity scores into biased training of recommender systems. 
Another recent work \cite{wang2022unbiased} analyzes the data bias mechanism in the sequential recommendation and re-weights the training parameters to reduce bias using inverse propensity scores.

\textbf{Backdoor Adjustment.}
Unlike the reweighting operation in IPS methods, \textit{backdoor adjustment} achieves fairness by blocking off relationships that lead to biases. When removing confounding factors, backdoor adjustment methods require the causal relationship between variables satisfies the backdoor criterion.
Here, the variable set $Z$ satisfies the backdoor criterion on a causal relationship $u \rightarrow v$ in the causal graph, indicating $Z$ satisfies (1) there is no descendant node of $u$ in $Z$, and (2) Every path between $u$ and $v$ that leads to $u$ is blocked by $Z$.
For example, as shown in Fig. \ref{causalgraph} (c), the history $H$ affects the representation of $u$ and $v$. 
To block the influence of $h$ on $u$ and $v$, backdoor adjustment methods employ a $do$ operation in cutting off the edge of $h \rightarrow u$ and the edge of $h \rightarrow v$, which can be formulated as $Pr(y|do(u),do(v)) = Pr (y|u,v).$

Although the backdoor adjustment is effective in removing unfairness, it faces efficiency challenges resulting from an unlimited sample space of confounding factors.
To this end, Wang et al. \cite{DBLP:journals/corr/abs-2105-10648} introduce DecRS, a model that uses data approximation and KL divergence to adjust the backdoor criterion. 
To lessen document-level label bias in text-contained recommender systems, DeSCoVeR \cite{rajanala2022descover} uses causal backdoor adjustment and sentence-level keyword bias elimination techniques in a semantic context. An inference model involving popularity-bias deconfounding and adjusting (PDA) is proposed by Zhang et al. \cite{zhang2021causal} as a new inference approach. It employs backdoor adjustment during model training to eliminate confusion caused by popularity bias.

\textbf{Counterfactual Inference.}
Counterfactual inference methods for fairness construct a counterfactual causal graph \cite{wei2021model} based on the real causal graph through some fairness-concerning actions (e.g., changing the values of some sensitive attributes like gender, age, race \cite{lesota2021analyzing,wu2021learning}). 
Recommender systems are fair and unbiased (i.e., counterfactual fairness in Section \ref{conceptsss}) if the recommendation results in the real world and the counterfactual world are the same. 
The intuition of counterfactual inference methods can be understood as ``if a fairness-concerning action (e.g., modifying the value of gender) cannot change recommendation results, then results from the recommender systems are not affected by the action-object (e.g., gender)." 
A representative counterfactual inference method for fairness is the framework called MACR \cite{wei2021model}, which is proposed to mitigate popularity bias.
MACR assumes that the recommendation interaction matrix $I_{uv}$ is affected by user $u$, item $v$, and the user-item matching ranking score $\hat{y}_{r}$, i.e., $y_{uv} = \hat{y}_{r}*\sigma(\hat{y}_{v})*\sigma(\hat{y}_{u})$, 
$\hat{y}_{v}$ indicates the influence from item popularity, and $\hat{y}_{u}$ represents the extent of the user $u$ interact with items. 
The higher value of the $\hat{y}_{u}$, the more likely the user is affected by the popularity of the item. 
Through counterfactual inference, the causal graph of the real world is transformed into the causal graph of the counterfactual world. 
The causal relation $I \rightarrow Y$ is removed by the following formula to alleviate the item popularity bias problem:
\begin{equation}\label{eq:ranking_test}
    \hat{y}_{r}*\sigma(\hat{y}_{i})*\sigma(\hat{y}_{u}) - c*\sigma(\hat{y}_{i})*\sigma(\hat{y}_{u}),
\end{equation}
here the hyperparameter $c$ controls the influence of user and item properties on the prediction result. The inference can be interpreted logically as a ranking adjustment based on $\hat{y}_{ui}$. 

FairTED \cite{acharyya2020fairyted} creates counterfactual samples of sensitive attributes to make sure that the speaker's sensitive attribute (i.e., gender) cannot influence the TED talk quality prediction. Specifically, when generating counterfactual samples, the score of presentations with female speakers are assigned as the score of presentations with the same contents and male speakers. These counterfactual samples are added into the training dataset to develop recommender systems with counterfactual fairness on gender. 
To mitigate the popularity bias and improve explainable fairness, Ge et al. \cite{DBLP:conf/sigir/GeTZXL0FGLZ22} propose a framework called CEF, which uses counterfactual inference (i.e., introducing small changes in the features) to find the root cause of the model’s bias. In CEF, the scores of each feature, which are calculated from counterfactual recommendation results, are regarded as fairness explanations.
Moreover, Li et al. \cite{10.1145/3404835.3462966} counterfactually infer that user-sensitive features should be orthogonal to user embeddings, and make fair personalized recommendations by removing user-sensitive features.
In the music streaming media recommender system, the user may have a situation where the music is playing incorrectly. Zhang et al. \cite{zhang2022counteracting} suggest a counterfactual learning strategy to correct user feedback that has been incorrectly categorized.

\subsubsection{Adversarial Learning for Fairness} 
\begin{figure*}[htpb]
	\centering
	\includegraphics[width=0.95\textwidth]{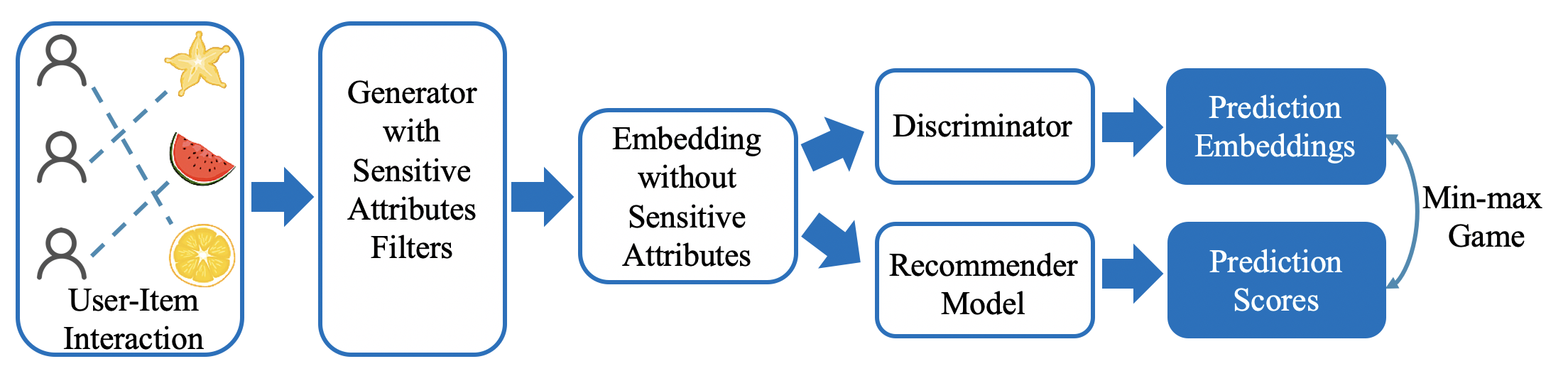} 
	\caption{Adversarial-learning-based fairness-aware methods. The adversarial-learning-based methods first use a generator and a sensitive attribute filter to remove sensitive information, resulting in embeddings without sensitive information, which are used for recommendation predictions. The corresponding sensitive attributes are predicted by the discriminator from the filtered embeddings. The generator and the discriminator engage in a max-min game.}
	\label{adv}
\end{figure*}

Adversarial learning is a method commonly used in recommender systems to remove sensitive attributes. An adversarial-learning-based fairness-aware framework generally consists of a generator that produces node embeddings and a discriminator that predicts sensitive features from these generator outputs. 
By playing a min-max game with the discriminator and generator, adversarial-learning methods are able to learn fair representations.  
Passing a negative gradient by predicting the sensitive attribute enables the model to fool the discriminator, so that the information content of the sensitive attribute is continuously reduced. 
When the discriminator cannot predict the sensitive feature value, the output of the generator is considered to be decoupled from the sensitive feature. 
The general form of loss for adversarial-learning-based fairness-aware methods can be formulated as:
\begin{equation}
    \mathcal{L} = \mathcal{L}_{per}(z,y) + \mathcal{L}_{adv}(y,s),
\end{equation}
where $z$ is a set of the generated representations of the generator, $y$ is a set of predictions, and $s$ is a set of the predictions of the discriminator.
Fig. \ref{adv} shows a general framework for adversarial-learning-based fairness-aware methods.
Recent work \cite{wei2022comprehensive} is a study addressing cold-start bias. To swiftly adjust to cold-start new users, it suggests a comprehensive fair meta-learning framework (CLOVER) to gather a general understanding of user preferences. CLOVER establishes that recommenders with ratings based on these representations will also fulfill counterfactual fairness if individual fair adversarial games converge to optimal solutions. Through adversarial learning, CLOVER improves the fairness of the cold-start problem in recommender systems.
Based on sensitive attributes, Wu et al. \cite{DBLP:conf/aaai/WuWWH021} craft a bias-aware user embedding, which is specially used to capture the bias. Furthermore, it also learns a bias-free user embedding, which is only used to encode attribute-independent information that the user interests. The fairness of the model is guaranteed by orthogonalizing the two types of embeddings.
Liu et al. \cite{DBLP:journals/kbs/LiuZZL000F22} propose an adversarial graph neural network (GNN) to prevent users from being affected by sensitive features of neighboring users. In particular, they propose two fairness constraints to address the failure and inefficiency of adversarial classifiers in the training data. 
Rus et al. \cite{DBLP:journals/corr/abs-2209-09592} use adversarial learning to mitigate gender bias in word embeddings obtained from recruitment-related text, which aims to provide unbiased job recommendations to job applicants. To alleviate multi-sided fairness. 
Liu et al. \cite{liu2022mitigating} design explicit and implicit adversarial fairness discriminators. Explicit discriminators aim to address biases from a local perspective, while implicit discriminators focus on addressing biases from a global perspective. The fairness generator and discriminator are trained adversarially together to mitigate bias.
In their paper, Wu et al. \cite{wu2022selective} propose a fairness awareness framework relying on prompts-based bias eliminators in combination with adversarial training.
Li et al. \cite{li2022fairgan} devise a generative adversarial network(GAN), named FairGAN, designed to generate negative signals of users to ensure data fairness. These signals enable FairGAN to complete the best item exposure ranking.

\subsubsection{Reinforcement Learning for Fairness}
\begin{figure*}[htpb]
	\centering
	\includegraphics[width=0.7\textwidth]{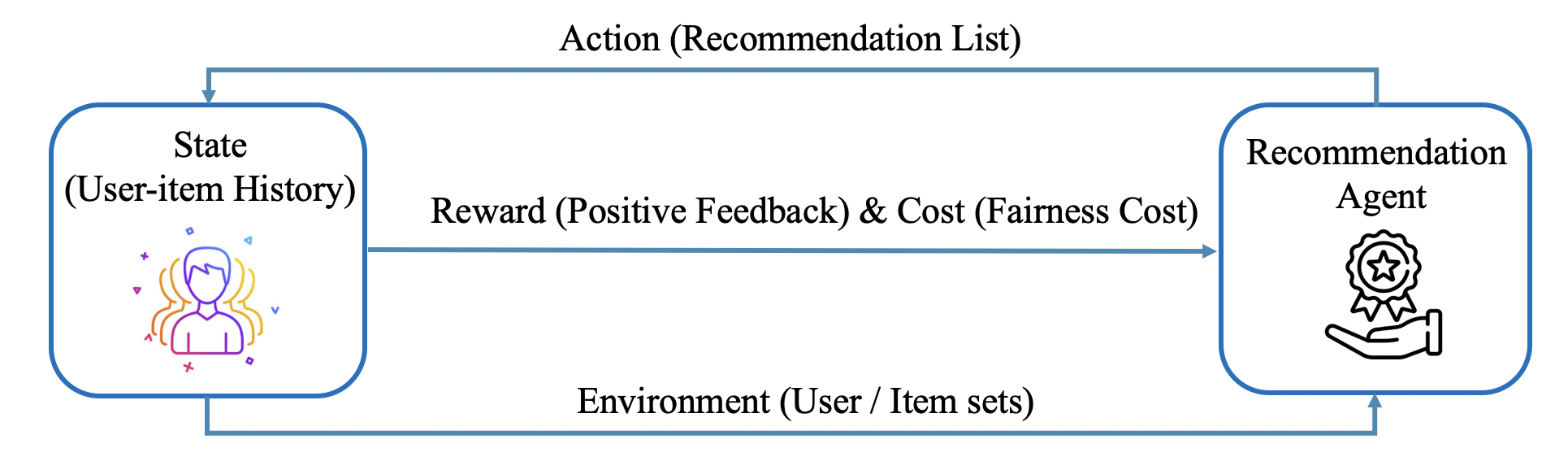} 
	\caption{Reinforcement-learning-based fairness-aware methods. A recommender system retains the learned experience in the process of interacting with the environment. When interacting in the next round, the behavior with the largest positive feedback and the smallest fairness cost will be selected.}
	\label{RL}
\end{figure*}

Recently, some studies take the dynamic interactions among users, items, and recommender systems into consideration, and model this feedback loop as a Markov Decision Process (MDP) \cite{DBLP:journals/corr/abs-2205-13619}. With this view, reinforcement learning (RL) methods are used in training recommendation strategies from users' historical information to learn their preferences. 
In this survey, \textit{reinforcement learning for fairness} indicates methods that introduce fairness-concerning feedback (e.g., dynamic fairness) during the training of RL-based recommender systems.
Fig. \ref{RL} shows a general framework for reinforcement-learning-based fairness-aware methods.

Long-term fairness and dynamic fairness are two typical concepts concerning mitigating time bias (see Section \ref{conceptsss}), which can be involved in the training of RL-based recommender systems. Here we introduce some representative reinforcement learning for fairness methods.
Ge et al. \cite{ge2021towards} majorly focus on the unfairness of item exposure across groups caused by time bias. According to their opinion, fairness constraints ought to change over time. 
For instance, an item may no longer be popular at time $t+n$, but if it is still exposed in accordance with the fairness requirements from time $t$ earlier, the long-term dynamic changes in fairness are disregarded. 
To enable the model to dynamically adjust the recommendation strategy and guarantee that the fairness requirements are always satisfied with environmental changes, they propose a fairness-constrained reinforcement learning recommender system, modeling the recommendation process as a constrained Markov Decision Process (CMDP). 
CMDP proposes two dynamic fairness constraints for reinforcement learning. The first is the population equality constraint, which requires equal average exposure for each group of items. This constraint is enforced at all reinforcement learning iterations. The second is the exact-k fairness constraint, which requires that the length of protected candidates in each recommendation list is statistically below a given threshold. 
In interactive recommender systems (IRS), Liu et al. \cite{liu2021balancing} dynamically maintain the long-term trade-off between accuracy and fairness by offering a method called FairRec. 
Using reinforcement learning, recommendations are generated by combining user preferences with system fairness in FairRec.
In addition, FairRec introduces a concept called weighted proportional fairness to ensure the fairness of item exposure.

Moreover, some approaches have proposed employing multi-objective reinforcement learning to improve fairness.
For instance, Ge et al. \cite{ge2022toward} investigate the Pareto optimal/effective fairness-utility trade-off problem in the recommendation process. Using multi-objective reinforcement learning, they suggest creating a fairness-aware recommendation framework (MoFIR), which introduces conditional networks and modifies the network according to user preferences. 
Fu et al. \cite{DBLP:conf/cikm/FuXGMZ21} propose a multi-objective MDP-based framework, namely Popcorn, for eliminating popularity bias in conversational recommender systems. Popcorn effectively balances recommendation performance and item popularity through a real-time semantic understanding of user history to avoid long-tail effects.

\subsubsection{Ranking Optimization for Fairness}
\begin{figure*}[htpb]
	\centering
	\includegraphics[width=0.7\textwidth]{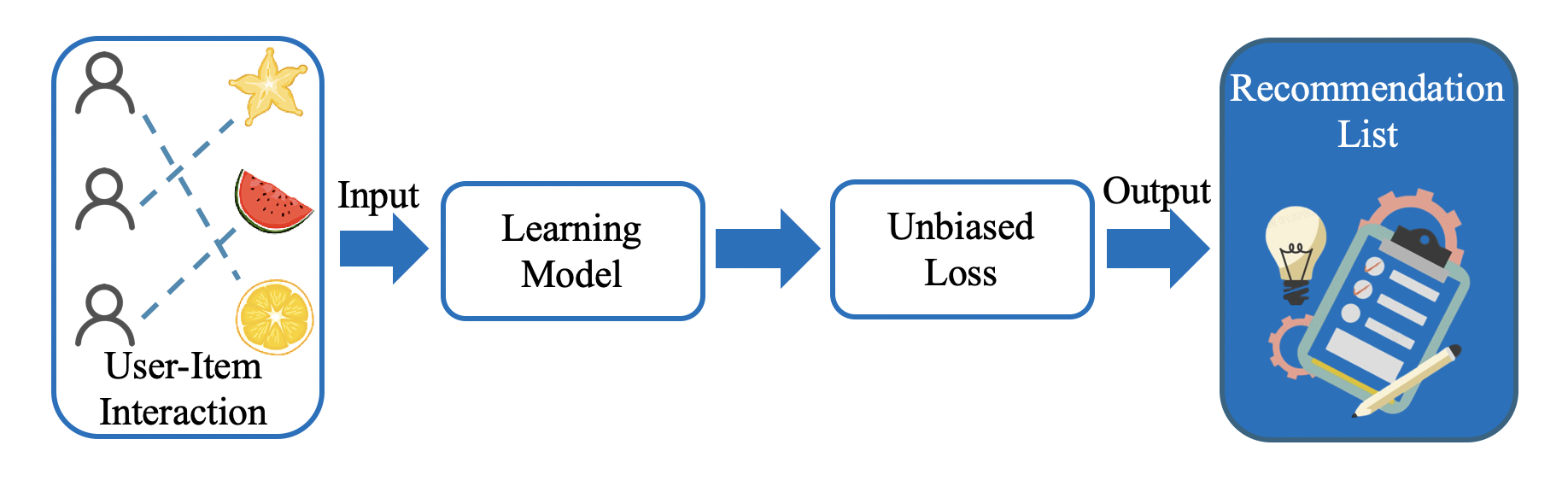} 
	\caption{Ranking-based fairness-aware methods. Such methods use a carefully designed unbiased loss for ranking.}
	\label{rank}
	\vspace{-1mm}
\end{figure*}

Ranking is an important part of the recommendation algorithm. The recommender system recommends items for a user according to the ranking of items. Specifically, the ranking algorithm sorts candidate items according to users' preference, and generates a ranking list. Top-scoring candidates receive the most exposure and are ranked first. The top-$k$ candidates are usually returned. A learnable recommender system usually uses a loss function for ranking. Depending on the flaw in the design, some biases, such as popular bias and exposure bias, can be amplified in losses with these flaws, which may lead to unfair outcomes. 
In this survey, \textit{ranking optimization for fairness} indicates methods that reduce unfairness in recommender systems by employing unbiased loss functions.
Fig. \ref{rank} shows a general framework for ranking-based fairness-aware methods.

There are two types of ranking losses commonly used in a learnable recommender system: point-wise loss and pair-wise loss. A point-wise loss, such as binary cross-entropy (BCE) \cite{DBLP:conf/aaai/DengHWLY19} and mean squared error (MSE) \cite{DBLP:conf/wsdm/0003ZS021}, minimizes the difference between the calculated recommendation score and the ground truth to capture user preferences for individual items. 
Pair-wise loss, such as Bayesian Personalized Ranking (BPR) \cite{DBLP:conf/uai/RendleFGS09}, maximizes the user preference gap between the observed items and the unobserved items.
The observed items are interacted with users and the unobserved items have no interaction with users. The BPR loss is a pair-wise ranking algorithm, that performs pairwise comparisons of items, and learns the order of relevant pairs from the comparisons to rank. The BPR loss encourages the observed item's prediction to be higher than the unobserved item's prediction. The BPR loss can be summarized as:
\begin{equation}\label{bpr}
    \mathcal{L}_{BPR}=-\sum_{(u,v_1,v_2)\in D_S} \ln \sigma(y_{u,v_1}-y_{u,v_2}),
\end{equation}
where $y_{u,v}$ is the predicted score, $D_S=\{(u,v_1,v_2) \mid I_{u,v_1}=1,\ I_{u,v_2}=0\}$, and $I_{u,v}$ is the interaction between the user $u$ and the item $v$.
$I_{u,v}=1$ indicates $v$ is an observed item, and vice versa.

It can be seen from the above equation that the BPR loss is affected by the observed item. Popular items tend to have higher observation than unpopular items, thus, they have a high exposure probability. The more items are observed by the user, the higher the exposure for the user, which leads to popularity bias and exposure bias. Flaws in the BPR loss cause the popularity bias to be amplified during training. Therefore, Wan et al. \cite{wan2022cross} propose a cross pairwise ranking loss $L_{CPR}$ for unbiased training:
\begin{equation}
    L_{CPR} = - \sum_{(u_1,u_2,v_1,v_2)\in D_c} ln \sigma[\frac{1}{2}(y_{u_1,v_1}+y_{u_2,v_2}-y_{u_1,v_2}-y_{u_2,v_1})], 
    \label{cpr}
\end{equation}
where $D_c=\{(u_1,u_2,v_1,v_2) \mid I_{u_1, v_1}=1,I_{u_2, v_2}=1,I_{u_1, v_2}=0,I_{u_2, v_1}=0\}$ denotes the training data, and $I$ is the user-item interaction set. The CPR loss uses cross pair-wise interactions as training samples. Given two users and their interacted items, the unobserved data is obtained by exchanging items for the user. CPR decomposes exposure probabilities of items into a user-specific, item-specific propensity and user-item relevance, which are not independent of each other and thus contain bias. 
These predicted scores expressed by exposure biases can cancel the exposure biases each other according to the equation \ref{cpr}, therefore the CPR loss is unbiased. There are also ranking methods that analyze the flaws of BPR loss \cite{DBLP:conf/sigir/ZhuWC20, DBLP:conf/kdd/BeutelCDQWWHZHC19}. However, these methods achieve unbias ranking by adding regularization terms, so we won't go into details here.

\subsubsection{Others}
In addition to the above-mentioned typical methods, there are also some niche methods to enhance the fairness of recommender systems.
Zhou et al. \cite {zhou2021contrastive} theoretically demonstrate that contrastive loss can replace the inverse propensity score to reduce exposure bias. 
Shen et al. \cite{DBLP:journals/corr/abs-2110-06475} learn cross-scenario user interests through an attention network, and propose a fairness factor to gauge how important each scenario is.
Zheng et al. \cite{zheng2022cbr} design a context-bias-aware recommendation model. They use attention networks to infer negative user preferences and eliminate contextual bias caused by the combined interaction between multiple items. 
Li et al. \cite{li2022fairsr} design an end-to-end model for time-influenced sequential recommendation by weighting the embeddings to mitigate the unfair distribution of user attributes over items. 

\subsection{Post-processing Methods for Fair Recommender Systems} 
\begin{figure*}[htpb]
	\centering
	\includegraphics[width=0.8\textwidth]{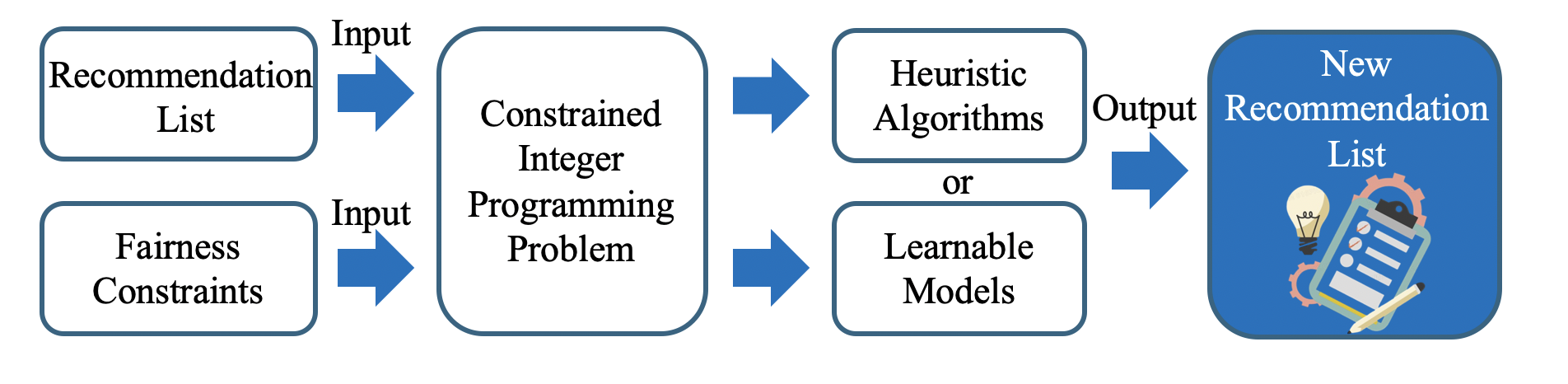} 
	\caption{The pipeline of re-ranking methods for fair recommender systems. These methods use recommendation lists and fairness constraints to construct integer programming problem, and use heuristic 
algorithms or learnable models to re-rank the recommendation lists.
	}
	\label{rerank}
\end{figure*}

From the view of fairness, the recommendation results from target systems are usually not optimal, since these results potentially don't take factors concerning fairness (e.g., interactions between items, manual intentions, and differences in user preferences) into consideration.
For enhancing fairness, \textit{post-processing} methods aim to rearrange the recommendation results provided by target models, which are treated as black-box during the rearrangement (i.e., re-ranking), after the training of recommender systems. 
Re-ranking methods include manual-based re-ranking and algorithmic-based re-ranking.
Here, we mainly introduce the re-ranking method that helps to boost the recommender system fairness, without repeating the manual re-ranking. 
Depending on whether or not the manual intervention is required, current re-ranking methods can be categorised into manual re-ranking and algorithmic-based re-ranking methods. 
In this survey, we focus on the algorithmic-based re-ranking methods and divide them into \textit{non-parametric re-ranking} and \textit{parametric re-ranking}, according to if a parameter learning process is required to conduct re-ranking.
Fig. \ref{rerank} shows a general framework of re-ranking methods for fair recommender systems.

\subsubsection{Non-parametric Re-ranking}
In this survey, \textit{non-parametric re-ranking} methods represent learning-free algorithms to conduct re-ranking concerning fairness. 
Heuristic methods used to achieve fairness are generally non-parametric re-ranking strategies, which treat the re-ranking process as an integer programming problem. Under specified fairness constraints, the optimal re-ranking outcomes are then found through heuristic search methods. Specifically, given original top-$k$ recommendation results for each user from recommender systems, heuristic methods are designed to maximize the total preference score with respect to fairness by rearranging original recommendations, which can be formulated as
\begin{equation}
    \begin{aligned}
         \arg &\max_{I_{uv}} \sum_{u=1}^n\sum_{v=1}^N I_{uv}S_{uv},\\
        s.t. & C(g_1, g_2)<\xi,\\
             & I_{uv}\in{[0,1]},
    \end{aligned}
\end{equation}
where $I_{uv}$ is an interaction matrix that reflects whether item $v$ should be recommended to user $u$,  $C(\cdot)$ represents a fairness constraint function, $g_1$ and $g_2$ are two different groups. The objective function is to maximize the recommendation score under fairness constraints, which is composed of user-item matching and user preference.
By treating the optimization as a 0-1 integer programming problem, a series of works employ heuristic algorithms to conduct re-ranking.
Here we present some representative methods.

To ensure fairness across various groups, Fu et al. \cite{fu2020fairness} utilized a heuristic re-ranking algorithm that constructs bias-constrained explainable recommendations on knowledge graphs. 
Similarly to this, Li et al. \cite{DBLP:conf/www/LiCFGZ21} proposed a similar approach, employing a heuristic-based method for re-ranking users, aiming to enhance group fairness considering their attributes.
According to Wu et al. \cite{DBLP:journals/corr/abs-2104-09024}, there's a model called TFROM, which is designed to enhance exposure fairness for both users and providers. In this model, the recommendation list's length is equated to the capacity of a knapsack, and the items symbolize the objects within this knapsack.
TFROM uses a heuristic search algorithm to solve the knapsack problem. 
Zhu et al. \cite{DBLP:conf/sigir/ZhuKNFC21} present a novel re-ranking framework including a score scaling model, which is used to re-rank the biased recommendation results.
Naghiaei et al. \cite{naghiaei2022cpfair} convert the recommendation optimization problem into a 0-1 integer programming and knapsack problem, and then propose employing greedy algorithms or solvers (e.g., Gurobi) to perform re-ranking.
Moreover, Mansoury et al. \cite{mansoury2020fairmatch} transform the item exposure inequity problem in popularity into a graph maximum flow problem to conduct re-ranking.
Recent work \cite{DBLP:conf/www/PatroBGGC20} design a two-stage approach to transform the problem into a max-min problem, using greedy strategies to ensure maximum exposure for producers and sorting to ensure fairness for each user.

\subsubsection{Parametric Re-ranking}
In this survey, all re-ranking methods that require a learning process to improve the fairness of recommender systems during the post-processing stage are called \textit{parametric re-ranking} methods. 
Existing study \cite{DBLP:conf/sigir/ZhuKNFC21} shows that, compared with non-parametric re-ranking methods, parametric re-ranking methods own better debiasing ability and are more competent when enhancing equality opportunity-based fairness. 
Here we introduce several typical parametric re-ranking methods.
Given the recommendation results from the target systems, Zhu et al. \cite{DBLP:conf/sigir/ZhuKNFC21} employ a learnable autoencoder module to enhance fairness and ensure the rearranged results have the same distribution as the original recommendations.



\section{Datasets and Evaluation for Fair RSs}
\label{dataset}
In this subsection, we provide a brief overview of the most commonly utilized datasets and assessment metrics in the context of fairness-aware recommender systems.

\subsection{Datasets}

\begin{table*}
\setlength{\abovecaptionskip}{0pt}
\setlength{\belowcaptionskip}{0pt}
\small
\caption{Summary of The Selected Datasets. We present the dataset, detailing its cardinality—specifically the number of users, items, and edges, its applicability to fairness issues, and associated reference relations.}

\begin{center} 
{

\begin{tabular}{lrrlll}
\toprule
Datasets                      & \#Users                 & \#Items                  & \#Edges                    & Fairness Issues & References                                                                                                                       \\ \midrule
\multirow{3}{*}{Amazon}       & \multirow{3}{*}{3,915}  & \multirow{3}{*}{2,549}   & \multirow{3}{*}{77,328}    & Exposure Bias   & \cite{DBLP:journals/corr/abs-2104-09024}                                                                                         \\ \cmidrule(l){5-6} 
                              &                         &                          &                            & Training Bias   & \cite{10.1145/3460231.3478858}                                                                                                   \\ \cmidrule(l){5-6} 
                              &                         &                          &                            & User Bias       & \cite{DBLP:conf/sigir/ZhuWC20}                                                                                                   \\ \midrule
\multirow{2}{*}{Beauty}       & \multirow{2}{*}{22,363} & \multirow{2}{*}{12,101}  & \multirow{2}{*}{198,502}   & Exposure Bias   & \cite{zhou2021contrastive}                                                                                                       \\ \cmidrule(l){5-6} 
                              &                         &                          &                            & User Bias       & \cite{fu2020fairness}, \cite{DBLP:conf/www/LiCFGZ21}, \cite{li2022fairgan}                                                       \\ \midrule
\multirow{3}{*}{Epinion}      & \multirow{3}{*}{49,290} & \multirow{3}{*}{139,738} & \multirow{3}{*}{664,828}   & Exposure Bias   & \cite{naghiaei2022cpfair}                                                                                                        \\ \cmidrule(l){5-6} 
                              &                         &                          &                            & Popularity Bias & \cite{zhu2022fighting}                                                                                                           \\ \cmidrule(l){5-6} 
                              &                         &                          &                            & User Bias       & \cite{10.1145/3437963.3441820}                                                                                                   \\ \midrule
\multirow{3}{*}{Lastfm}       & \multirow{3}{*}{1,892}  & \multirow{3}{*}{17,632}  & \multirow{3}{*}{92,834}    & Exposure Bias   & \cite{naghiaei2022cpfair}, \cite{wu2021learning}, \cite{DBLP:conf/www/PatroBGGC20}, \cite{DBLP:journals/tois/MansouryAPMB22}     \\ \cmidrule(l){5-6} 
                              &                         &                          &                            & Training Bias   & \cite{10.1145/3460231.3478858}                                                                                                   \\ \cmidrule(l){5-6} 
                              &                         &                          &                            & User Bias       & \cite{DBLP:conf/aaai/DoCAU22}                                                                                                    \\ \midrule
\multirow{3}{*}{MovieLens}    & \multirow{3}{*}{943}    & \multirow{3}{*}{1,349}   & \multirow{3}{*}{99,287}    & Exposure Bias   & \cite{naghiaei2022cpfair}, \cite{DBLP:journals/tois/MansouryAPMB22}                                                              \\ \cmidrule(l){5-6} 
                              &                         &                          &                            & Popularity Bias & \cite{DBLP:conf/cikm/FuXGMZ21}, \cite{huang2022different}                                                                        \\ \cmidrule(l){5-6} 
                              &                         &                          &                            & User Bias       & \cite{10.1145/3404835.3462966}, \cite{sato2022enumerating}                                                                       \\ \midrule
\multirow{6}{*}{MovieLens 1M} & \multirow{6}{*}{6,040}  & \multirow{6}{*}{3,706}   & \multirow{6}{*}{1,000,209} & Cold-start Bias & \cite{DBLP:conf/sigir/ZhuKNFC21}, \cite{wei2022comprehensive}                                                                    \\ \cmidrule(l){5-6} 
                              &                         &                          &                            & Exposure Bias   & \cite{wu2022joint}, \cite{DBLP:journals/corr/abs-2005-12964}, \cite{wu2021learning}                                              \\ \cmidrule(l){5-6} 
                              &                         &                          &                            & Time Bias       & \cite{ge2021towards}, \cite{li2022fairsr}                                                                                        \\ \cmidrule(l){5-6} 
                              &                         &                          &                            & Training Bias   & \cite{10.1145/3460231.3478858}                                                                                                   \\ \cmidrule(l){5-6} 
                              &                         &                          &                            & Popularity Bias & \cite{10.1145/3437963.3441820}, \cite{mena2021popularity}, \cite{abdollahpouri2020connection}                                    \\ \cmidrule(l){5-6} 
                              &                         &                          &                            & User Bias       & \cite{DBLP:conf/sigir/ZhuWC20}, \cite{DBLP:journals/corr/abs-2105-10648}, \cite{zhu2022fighting}, \cite{10.1145/3442381.3449904} \\ \midrule
\multirow{3}{*}{Yahoo!R3}     & \multirow{3}{*}{5,400}  & \multirow{3}{*}{1,000}   & \multirow{3}{*}{183,179}   & Popularity Bias & \cite{mena2021popularity}, \cite{10.1145/3460231.3474263}, \cite{abdollahpouri2020connection}                                    \\ \cmidrule(l){5-6} 
                              &                         &                          &                            & Training Bias   & \cite{10.1145/3460231.3478858}                                                                                                   \\ \cmidrule(l){5-6} 
                              &                         &                          &                            & User Bias       & \cite{saito2020unbiased}, \cite{liu2022rating}, \cite{10.1145/3383313.3412252}                                                   \\ \midrule
\multirow{4}{*}{Yelp}         & \multirow{4}{*}{25,677} & \multirow{4}{*}{25,815}  & \multirow{4}{*}{731,671}   & Exposure Bias   & \cite{DBLP:conf/sigir/GeTZXL0FGLZ22}                                                                                             \\ \cmidrule(l){5-6} 
                              &                         &                          &                            & Popularity Bias & \cite{wei2021model}, \cite{DBLP:conf/cikm/FuXGMZ21}                                                                              \\ \cmidrule(l){5-6} 
                              &                         &                          &                            & Training Bias   & \cite{10.1145/3460231.3478858}                                                                                                   \\ \cmidrule(l){5-6} 
                              &                         &                          &                            & User Bias       & \cite{DBLP:conf/sigir/ZhuWC20}, \cite{zhu2022fighting}, \cite{DBLP:conf/ijcai/YangWJ0PC22}                                       \\ \bottomrule
\end{tabular}

}\label{data}
\end{center}
\vspace{-0.5cm}
\end{table*}

In this section, we will present several datasets frequently employed in experiments related to the credibility of recommendation systems. Due to the complexity of fairness study in recommender systems, one specific dataset can be applied in addressing different bias issues. 
For example, the Amazon Review dataset is used in mitigating user attribute bias, model weights bias, and popularity bias. 
It is beneficial to use a variety of datasets to detect unfairness in recommender systems and evaluate fairness-enhancing methods. Here we summarise commonly used datasets in Table \ref{data}, demonstrating their basic information, 
and possible scenarios (e.g., exposure bias, and what kind of recommender systems can be used).

In addition to the datasets mentioned in Table \ref{data}, some other datasets can also be used to mitigate unfairness. 
For example, Gowalla \cite{yang2022trading}, Ctrip \cite{DBLP:journals/corr/abs-2104-09024} can be used to solve the Pairwise bias. CIKM and AliEc can be used to solve the user attribute biases \cite{wu2022selective}. MinD \cite{qi2022profairrec} and COCO \cite{Mooc} can be used to solve user behavior biases. CiteULike and XING \cite{DBLP:conf/sigir/ZhuKNFC21} can be used to solve cold-start bias. Clothing, CellPhones \cite{fu2020fairness} are used to solve Group fairness. Ciao \cite{wei2021model} can be used to solve Feedback bias.

\subsection{Evaluation}
Evaluating a fairness-aware recommender system can be measured in terms of its accuracy and fairness. 
To avoid bias in recommendation results, this section presents accuracy metrics that won't harm fairness first, then metrics for assessing the fairness of recommender systems.

\textbf{Accuracy Metrics for Fairness.}
There are some evaluation metrics used for evaluating the quality of a model. The common evaluation metrics are as follows:
\begin{itemize}
    \item\textit{Area Under Curve (AUC)} \cite{DBLP:conf/www/SunPZF21} is the region contained within the Receiver Operating Characteristic (ROC) curve and the coordinate axis. The maximum possible value of AUC is one. The closer the AUC is to 1, the higher the authenticity of the detection method; and vice versa.
    \item \textit{Precision (P)} \cite{DBLP:conf/recsys/AntogniniF21} is the proportion of correctly classified positive examples to all positive examples that are classified, and Precision@k (P@k) represents the Precision of the top-$k$ items in the list.

    \item \textit{Recall (R)} \cite{DBLP:conf/sigir/ZouKRRSL22} refers to how many of the positive samples are successfully found by the model. Recall@k indicates the recall of the first $k$ items in the result list.

    \item \textit{Average Precision (AP)} \cite{krichene2022sampled} calculates the average precision. The higher the value of AP@K, the more relevant items are present in the top$-K$ recommendations and the higher these relevant items are ranked.
    \item \textit{Normalize Discounted Cumulative Gain (NDCG@k)} \cite{DBLP:conf/colt/WangWLHL13} provides different degrees of relevance, ranking the results according to relevance and weighting them uniformly.
\end{itemize}

\textbf{Metrics to Evaluate Fairness.}
There are some commonly used metrics for measuring the fairness of recommender systems:
\begin{itemize}
    \item \textit{Gini coefficient} is an economic indicator that is used to measure the income inequality of a country or region \cite{dorfman1979formula}. It is also often used to measure the fairness of recommender systems, which measures the value inequality of a frequency distribution \cite{ge2021towards}. The value of Gini coefficient is between 0 and 1, and smaller values represent greater fairness. Its definition is:
\begin{equation}\label{eq:gini}
    Gini\ coefficient = \frac{1}{2n^2\overline{ev}}\sum_{i=1}^{n}\sum_{j=1}^{n} |ev_{i} - ev_{j}|,
\end{equation}
where $n$ indicates items' number, $ev_i$ represents the exposure frequency of item $v_i$, and $\bar{ev}$ donates the average of items' exposure frequency.
\item \textit{KL-divergence} computes the divergence between two distributions \cite{DBLP:journals/eswa/SilvaMD21}, which can be used to measure various biases.
Taking popularity bias as an example, KL-divergence can measure the distribution difference between user history records and popularity in recommendation results \cite{lesota2021analyzing}. 
The recommendations are fairer when the value of KL-divergence is lower. 
The expression of KL-divergence is
\begin{equation}
    D_{KL}(D_1(v),D_2(v)) = \sum_{v\in V}D_1(v)log\frac{D_1(v)}{D_2(v)},
\end{equation}
where $D_1(\cdot)$ refers to a distribution of item $v$ and $D_2(\cdot)$ refers to a fair distribution of item $v$.

\item \textit{Difference} mainly considers that if the distance between the two recommendation results is less than the fairness constraint coefficient, then the recommendation system is considered to be fair \cite{DBLP:conf/www/LiCFGZ21}, i.e.,
\begin{equation}
    |y_{v_i}-y_{v_j}|<\xi,
\end{equation}
where $y_{v_i}$ is a recommender system prediction of item $v_i$, $\xi$ is strictness parameter of fairness requirements. If the prediction score of $v_i$ and $v_j$ is smaller than the fairness constraint coefficient, we consider that the recommender system is fair.

\end{itemize}

In addition to the above metrics, we can also employ other metrics to evaluate fairness-aware recommender systems.
For example, Lesota et al. \cite{lesota2021analyzing} propose measures to account for prevalence bias from the median, various statistical moments, and measures of similarity that consider the entire prevalence distribution, including Mean, Median, Variance, Skew, Kurtosis, Kendall's $\tau$  rank-order correlation. 
If a recommender system offers suggestions that suit the preferences of a certain group of users, but fails to accurately reflect the preferences of another user group, it can be deemed unfair. 
An analytical metric called miscalibration is used by Abdollahpouri et al. \cite{abdollahpouri2020connection} to measure the degree to which a recommender system responds to the true preferences of the user. 
If the Hellinger distance \cite{abdollahpouri2020connection} between two distributions exceeds a threshold, then it is called miscalibration. According to Fu et al. \cite{DBLP:conf/cikm/FuXGMZ21}, two measures are used to measure popularity bias in conversational recommender systems (CRSs): personalized average recommendation popularity (PARP) and popularity-rank correlation for users (PRU). Mena-Maldonado et al. \cite {mena2021popularity} present two novel metrics for describing the relationship between prevalence and relevance distributions. The article experimentally identifies scenarios for the use of true-positive metrics or false-positive metrics. 
Ge et al. \cite{ge2021towards} propose a fairness indicator known as the popularity rate, which represents the proportion of popular items in a recommended list in relation to the total number of recommended items.

\section{Industrial Applications of Fair RSs}
\label{application}
\begin{table*}[ht!]
\small
\caption{Industrial Applications of Fairness-aware Recommender Systems. We delineate the applications by outlining their respective domains of existence, the prevalent bias that they grapple with, and the methodological approaches they adopt to counteract these biases.}

\begin{tabular}{@{}cclll@{}}
\toprule
\multicolumn{1}{l}{Domains}       & \multicolumn{1}{l}{Sub-domains}                                                           & Bias            & Methods                 & Reference                                            \\ \midrule
\multirow{2}{*}{E-commerce}       & \multicolumn{1}{l}{}                                                                      & Exposure Bias   & Ranking                 & \cite{DBLP:journals/corr/abs-2110-06475}             \\ \cmidrule(l){3-5} 
                                  & \multicolumn{1}{l}{}                                                                      & User Bias       & Re-ranking              & \cite{jingdong}                                      \\ \midrule
\multirow{2}{*}{Education}        & \multicolumn{1}{l}{}                                                                      & Exposure Bias   & Re-ranking              & \cite {gomez2022enabling}, \cite{marras2021equality} \\ \cmidrule(l){3-5} 
                                  & \multicolumn{1}{l}{}                                                                      & Popularity Bias & Ranking                 & \cite{boratto2019effect}                             \\ \midrule
\multirow{9}{*}{Social Activities} & \multirow{2}{*}{\begin{tabular}[c]{@{}c@{}}Job\\ Recommendation\end{tabular}}             & Exposure Bias   & Counterfactual Learning & \cite{10.1145/3357384.3358131}                       \\ \cmidrule(l){3-5} 
                                  &                                                                                           & User Bias       & Adversarial Learning    & \cite{DBLP:journals/corr/abs-2209-09592}             \\ \cmidrule(l){2-5} 
                                  & \multirow{3}{*}{\begin{tabular}[c]{@{}c@{}}News \\ Recommendation\end{tabular}}           & User Bias       & Regularization           & \cite{qi2022profairrec}                              \\ \cmidrule(l){3-5} 
                                  &                                                                                           & User Bias       & Adversarial Learning    & \cite {zheng2022cbr}                                 \\ \cmidrule(l){3-5} 
                                  &                                                                                           & Popularity Bias & Adversarial Learning    & \cite{qi2021pp}                                      \\ \cmidrule(l){2-5} 
                                  & \multirow{4}{*}{\begin{tabular}[c]{@{}c@{}}Streaming Media\\ Recommendation\end{tabular}} & Popularity Bias & Ranking                 & \cite{kirdemir2021assessing}                         \\ \cmidrule(l){3-5} 
                                  &                                                                                           & Time Bias       & Backdoor Adjustment     & \cite{zhan2022deconfounding}                         \\ \cmidrule(l){3-5} 
                                  &                                                                                           & Popularity Bias & Re-ranking              & \cite{Linkin}                                        \\ \cmidrule(l){3-5} 
                                  &                                                                                           & User Bias       & Adversarial Learning    & \cite{wu2022fairrank}                                \\ \bottomrule
\end{tabular}
\label{apps}
\end{table*}

This session focuses primarily on the application of fairness-aware recommender systems to real-world situations. We introduce the impact of fairness-aware recommender systems on life from application scenarios such as e-commerce and finance. Table \ref{apps} highlights real-world applications of fair recommender systems.

\subsection{Fairness in E-commerce Recommendations}


There are many applications of the fair correction recommendation system in the area of e-commerce. To address the uneven data distribution across scenarios and the systematic bias of disadvantaged items that emerge in the re-ranking phase, 
Shen et al. \cite{DBLP:journals/corr/abs-2110-06475} have introduced a solution to the aforementioned challenges in the form of a Scenario-Aware Ranking Network (SAR-Net). The SAR-Net is designed to capture a user's interests across various scenarios using two tailored attention modules. By assessing the importance of individual samples and adjusting predictions accordingly, the SAR-Net aims to mitigate data bias arising from human intervention.
Wu et al. \cite{jingdong} introduce that in e-commerce websites such as Amazon and JD.com, a feeds-based recommendation has become a mainstream recommendation mode, and users can scroll down to view more products recommended by feeds. In the rolling feeds recommendation, four grids with pictures are usually played on the same mobile phone screen, and the similarity of the four products will affect the user's judgment, thus introducing bias. The article considers this phenomenon to be a contextual bias. The authors propose an unbiased counterfactual learning method to eliminate contextual bias. The proposed method is applied to a real-world e-commerce website, JD.com.


\subsection{Fairness in Education Recommendations}
Users are hoped to receive an education without bias in a fair education recommender system.
G\'{o}mez et al. \cite {gomez2022enabling} believe that the geographic location of teachers has a strong impact on visibility and exposure. The re-ranking approach overcomes these phenomena by ensuring that each group receives the exposure expected, thereby ensuring that different providers are treated fairly. 
Boratto et al. \cite{boratto2019effect} explore how recommender systems in the context of popularity-biased massively open online courses. A comparison is made of existing algorithms relating to the popularity of courses, catalog coverage, and popularity of course categories. Marras et al. \cite{marras2021equality} provide a formal definition of the online education recommendation principle and propose a novel re-ranking method that is conscious of fairness, in an effort to strike a balance between personalization and recommendation opportunities.


\subsection{Fairness in Social Activities}
\subsubsection{Job Recommendations}
Job recommendation systems match job seekers with job information, and recommend job listings that meet job seekers' wishes, or lists of talented candidates that meet the requirements of the recruiter \cite{DBLP:conf/rif/Brek022}. 
However, user-sensitive attributes may cause discrimination for users in job recommender systems and reduce users' trust in them. Several researches are devoted to mitigating discrimination and improving the fairness of job recommendation systems.
Chen et al. \cite{10.1145/3357384.3358131} find that click-through rates for job advertisements decreased over time. Thus, they adopt the inverse propensity weighting method and customize a new loss function to rank the deviation of ad exposure position.
In a recent paper, Rus et al. \cite{DBLP:journals/corr/abs-2209-09592} demonstrate that gender bias can be removed from $12$ million job openings and $0.9$ million resumes through the use of a generative adversarial network, providing fair job recommendations to mitigate the pay gaps between different genders.

\subsubsection{News Recommendations}
News recommendation systems mainly recommend news to users on digital news sites. Qi et al. \cite{qi2022profairrec}  present ProFairRec, a news recommendation framework that prioritizes provider fairness. By integrating adversarial learning, the framework ensures that representations of fair news from providers remain unbiased during the recommendation process. They suggest the use of orthogonal regularization of provider-fair and biased representations to decrease the bias associated with news providers. Zheng et al. \cite{zheng2022cbr} argue that the contextual bias among news items may not be fully captured due to interactions among multiple items. To address this, they propose a novel context-bias-aware recommendation model aimed at eliminating context bias and achieving fairness in recommendations. 
Qi et al. \cite{qi2021pp} propose user encoders with popularity awareness to eliminate popularity bias from user behavior and achieve accurate interest modeling. In news recommendation, several recommender systems utilize multiple heads to capture correlations between news items based on representations from the news that users view.
Yi et al. \cite{yi2021debiasedrec} argue that news click behavior may also be biased by the way news is presented on online platforms. So this paper proposes a bias-aware personalized news recommendation approach called DebiasRec. Debiasrec trains a biased news recommendation model from biased click behavior and inferring the biased interests of users from the clicked news articles.

\subsubsection{Streaming Media Recommendations}
Streaming media recommendation systems include music recommendation, video recommendation, etc. 
As a means of increasing transparency and fairness to artists in music recommendation systems, Kirdemir et al. \cite{kirdemir2021assessing} find the presence of video recommendations in YouTube's structural and systematic biases in YouTube. By employing a graphical probabilistic approach, this study evaluates the structural properties of video recommendations. To eliminate undesired temporal bias, Zhan et al. \cite{zhan2022deconfounding} propose a duration-fault quantization (D2Q)-based watch-time prediction framework that allows for industrial production systems for scaling. The framework has been implemented within the Kuaishou App, a commercial video streaming platform. This has resulted in a substantial enhancement in predicting real-time video viewing time, thereby significantly improving real-time video consumption. 
A study by Shakespeare et al. \cite{shakespeare2020exploring} examines whether state-of-the-art collaborative filtering algorithms exacerbate or ameliorate artist gender biases. This work designs two methods for determining why differences are attributed to changes in the distribution of inputs based on gender and user preferences. 
Melchiorre et al. \cite{10.1145/3383313.3412223} construct a dataset comprising information about the music consumption habits and personality traits of Twitter users. This work analyzes the recommendation algorithms SLIM and EASE Mult-VAE. Their research results reveal notable differences in performance between user groups scoring high and low on certain personality traits. The recent work \cite{Linkin} proposes a fairness-conscious re-ranking framework for quantifying and mitigating algorithmic bias due to data bias. In an online A/B test of representative rankings of LinkedIn Talent Search or recommendations, the authors propose a strategy aimed at distributing ranking outcomes according to one or more safeguarded attributes, with the goal of achieving fairness principles like equal opportunity and population parity.  This large-scale deployment of a framework that deploys LinkedIn Recruiter to ensure fairness in the recruitment space without impacting business metrics has the potential to positively impact over 630 million LinkedIn members. Wu et al. \cite{wu2022fairrank} use adversarial learning to reduce bias arising from user-sensitive attributes. Furthermore, they utilize KL divergence to capture less candidate-aware bias.

\section{Connections with other Trustworthy Dimensions}
\label{connection}

Recommendation systems are one of the most crucial parts of the lives of people today. However, some recommendations lack moral basis and restraints, which undermines user confidence and possibly transgresses the law. 
A crisis of trust in the recommendation system can be sparked by a significant volume of biased training data or biased recommendation algorithms. 
Thus, it is uttermost important to have trustworthy recommender systems.
Ideally, a trustworthy recommender system should be open and transparent, and its methods for obtaining results should be explainable. 
The trustworthiness of a recommender system is mainly evaluated based on four ethical principles, including explainability, robustness, privacy, and fairness. Here we introduce other ethical principles beyond fairness.
\begin{itemize}
    \item \textbf{Explainability.} Explainability requires that the decisions of the recommender system should be understandable by people. Specifically, it requires that the decision-making process, and input and output relationships of recommender systems should be logically explained. However, most of the current recommendation models operate in the form of "black boxes," sometimes it is not always possible to explain why a recommender system produces a particular output or decision, which can lead to users' distrust of recommendation models. Therefore, explainability is crucial for building user trust in recommender systems.
    \item \textbf{Robustness.} The robustness of recommender systems refers to their ability to continue to operate normally when threatened or attacked. It requires the recommender system to be safe, reliable, and robust enough to handle errors or inconsistencies in all life cycle stages of the recommender system. Recommendation systems generally rely on user history records to build algorithm models. User history records contain a lot of junk data generated by systems and humans,  which may lead to a decrease in the fairness and accuracy of the model. Thus, a trustworthy recommendation algorithm must be robust.
    \item \textbf{Privacy.} Personal data collected by a recommender system should be safe and able to protect personal privacy. Private data and privacy must be protected throughout the life cycle of the recommender system. Private data encompasses both the information shared by the user and the information derived from the user's interactions with the system. A trustworthy recommendation system should take responsibility for preventing unlawful and unfair discrimination against users as a result of the collected data.
\end{itemize}

\begin{table*}
\setlength{\abovecaptionskip}{0pt}
\setlength{\belowcaptionskip}{0pt}
\small
\caption{Connections between Fair-aware RSs to Trustworthy RSs. We present various categories of fairness recommendation methodologies associated with trustworthiness properties, along with a succinct summary of their content.}

\scalebox{0.99} 
{

\begin{tabular}{@{}ccll@{}}
\toprule
Trustworthiness                 & \begin{tabular}[c]{@{}l@{}}Methods for\\ Fair RSs\end{tabular}              & Ref.                                     & Sketches                                                                                                                                                                                            \\ \midrule
\multirow{6}{*}{Explainability} & \multirow{4}{*}{\begin{tabular}[c]{@{}c@{}}Causal\\ Inference\end{tabular}} & \cite{DBLP:journals/corr/abs-2006-16977} & Causal analysis on the relationship between users' past and future behaviors.                                                                                                                       \\ \cmidrule(l){3-4} 
                                &                                                                             & \cite{DBLP:conf/sigir/GeTZXL0FGLZ22}     & \begin{tabular}[c]{@{}l@{}}An explainable weighting method is employed to rank counterfactual \\ recommendation outcomes effectively.\end{tabular}                                                \\ \cmidrule(l){3-4} 
                                &                                                                             & \cite{DBLP:conf/recsys/CornacchiaNR21}   & \begin{tabular}[c]{@{}l@{}}A counterfactual analysis and explanation are provided to bolster \\ the effectiveness of explanations and promote fairness in the process.\end{tabular}                 \\ \cmidrule(l){3-4} 
                                &                                                                             & \cite{fu2020fairness}                    & \begin{tabular}[c]{@{}l@{}}Presenting a fairness-constrained method that utilizes heuristic re-ranking to\\ address the issue of unfairness recommendations based on knowledge graphs.\end{tabular} \\ \cmidrule(l){2-4} 
                                & \begin{tabular}[c]{@{}c@{}}Data\\ Modification\end{tabular}                 & \cite{DBLP:journals/corr/abs-2006-16977} & \begin{tabular}[c]{@{}l@{}}Utilizing an inverse propensity score helps eliminate polarity bias in group\\ recommendations, ensuring a more robustness and fairness outcome.\end{tabular}            \\ \cmidrule(l){2-4} 
                                & Regularization                                                              & \cite{zhu2022cali3f}                     & Improving robustness and control fairness through L2 regularization loss.                                                                                                                           \\ \midrule
\multirow{2}{*}{Privacy}        & Causal Inference                                                            & \cite{wei2022comprehensive}              & Training models on risk-free, user-approved privacy data.                                                                                                                                           \\ \cmidrule(l){2-4} 
                                & Others                                                                      & \cite{wu2021learning}                    & Preventing the exposure of sensitive information within the learned embeddings.                                                                                                                     \\ \midrule
Discrimination                  & Re-ranking                                                                  & \cite{Mooc}                              & \begin{tabular}[c]{@{}l@{}}Utilizing re-ranking methods helps minimize discrimination,  \\ promoting fairness and equal representation in the results.\end{tabular}                                 \\ \midrule
\multirow{2}{*}{Diversity}      & Re-ranking                                                                  & \cite{DBLP:journals/tois/MansouryAPMB22} & Transforming the issue into a maximum flow problem to improve the diversity.                                                                                                                        \\ \cmidrule(l){2-4} 
                                & Regularization                                                              & \cite{DBLP:conf/um/SacharidisMW20}       & \begin{tabular}[c]{@{}l@{}}Incorporating regularization terms can enhance fairness and diversity,  \\ ensuring a more balanced and inclusive outcome.\end{tabular}                                  \\ \bottomrule
\end{tabular}

}

\label{connections}

\vspace{-0.5cm}
\end{table*}


A recommender system may contain a variety of untrustworthy issues, and biases that cause unfairness may also cause other untrustworthiness. For example, privacy and robustness issues caused by data bias (age and gender). The issue of fairness may arise simultaneously with other trustworthy issues in a recommender system. We are concerned with the fairness of the recommender system and introduce other trustworthy ethical principles based on the fairness content. We describe the connection between fairness and each trustworthy property below. Table \ref{connections} outlines the connections between fairness and other trustworthiness properties.

\subsection{Connections with Explainability}
Some methods for mitigating fairness issues can also be added to the explainability of models. For example, methods based on causal inference analyze the causes of bias and provide explanations for the decision-making process of recommendation models.
Xu et al. \cite{DBLP:journals/corr/abs-2006-16977} argue that the explainability of recommender systems involves causal analysis between the previous and future behaviors of users, which is bound to answer counterfactual questions. An example of the question can be ``What would happen if a different set of items were purchased." 
Counterfactual inference provides a fair framework for recommender systems, where the constructed counterfactual world explains why the model makes the output decisions. 
Therefore, counterfactual reasoning can simultaneously promote the explainability and fairness of recommender systems.
Ge et al. \cite{DBLP:conf/sigir/GeTZXL0FGLZ22} propose a counterfactual explainable fairness framework for group fairness. Specifically, they propose an explainable weighting method to rank the counterfactual recommendation results, which can be seen as an explanation for the final recommendation. 
Cornacchia et al. \cite{DBLP:conf/recsys/CornacchiaNR21} propose a model that fuses natural language processing and counterfactual inferencing to provide recommendations for the loans domain. This model provides users with fairness and transparent advice. 
The path-based method is a common method to make improvements to recommender systems' explainability. Fu et al. \cite{fu2020fairness} make improvements on user-item path distribution and fairness of the recommender system by designing a fairness-aware ranking algorithm.


\subsection{Connections with Robustness}
Robustness can be reflected in the ability to defense attacks on data and models. Data bias cause unfair, it is also vulnerable to attack, which makes the model less robust.
Fang et al. \cite{DBLP:conf/sacmat/Fang0MS22} propose to construct antidote data that mimics the rating behavior of users to mitigate data bias. These data are not considered anomalous data for attacking, thus it can make an improvement on the  model's robustness. The same method also be adopted by Rastegarpanah et al. \cite{DBLP:conf/wsdm/RastegarpanahGC19}.
Dokoupil et al. \cite{dokoupil2022robustness} argue that recommender systems should utilize as much unbiased data as possible, whereas real-world training data is biased. In this case, to make the recommender system robust, they use an inverse propensity score to remove polarity bias in group recommendations. As a result, the fairness of group recommendations has been improved. 
Zhu et al. \cite{zhu2022cali3f} design a local model and a global model to improve robustness and control fairness through l2 regularization loss. They improve model robustness and fairness through continuous gradient optimization.

%
\subsection{Connections with Privacy}
Training data in recommender systems may contain some user-sensitive attributes that may be considered private by users. Even if the user's sensitive data is well protected, privacy leakage may occur during the interaction with the recommender system. The leaked private information can be maliciously obtained by other users, which brings data bias to the recommendation system and causes unfairness.
Some works \cite{wei2022comprehensive, do2022online} establish a connection between fairness and privacy. Recommender systems may be unfair if users' private information is used extensively for personalization and when protected private data attributes like gender and ethnicity are misused. A bipartite graph is typically created organically by users and items in recommender systems. Directly abusing some user-item representations will nevertheless cause the leakage of user-sensitive information, even if user-item interactions don't contain any user-sensitive data. This is because user behavior and attributes are found to be correlated in social theory. For example, a person's privacy (such as his gender) can be inferred from his actions. Each user's embedding has a hidden connection to the behavior of similar users and users who have the same items, in addition to being tied to the user's behavior. 

Similarly, Wu et al. \cite{wu2021learning} make an effort to keep user privacy and sensitive information hidden from the recommender system. They transform the fairness-aware recommendation problem into learning fair user and item representations, and provide a GNN method (FairGo) to avoid any sensitive information from being revealed from the learned embeddings. To make a fair recommendation and prevent the spread of high-level sensitive information, FairGo designs an ego network, which is user-centric and links the purchased products of the user and the item. Fairgo designs an aggregation algorithm that prevents high-order information propagation in the ego network and achieves representation fairness. The ego network embedding and user-item embedding are mapped into the same space after learning the embedding. In this space, filters of sensitive information are used for filtering, and finally, fairness training is performed through graph adversarial learning. In a cold-start situation, where user-item interaction data is lacking, recommender systems leverage the data trained from non-cold-start scenarios as the proxy for cold-start user-item information. However, this approach can leak the privacy of non-cold-started user items. Wei et al. \cite{wei2022comprehensive} suggest training risk-free, user-approved private data, and then making privacy-preserving fair recommendations to cold-start users. 


\begin{table*}
\setlength{\abovecaptionskip}{0pt}
\setlength{\belowcaptionskip}{0pt}
\small
\caption{Future Directions for Fairness-aware Recommender Systems. We categorize future directions by segmenting them into distinct trajectories, elucidating the current deficiencies inherent in the fairness-aware RSs, the potential challenges to be encountered, and elucidating the advantages of resolving these issues.
}

\scalebox{0.99} 
{
\begin{tabular}{llll}
\toprule
\bf Directions & \bf Current Shortcomings &\bf Future Challenges & \bf Challenges-solving Benefits 
\\

\midrule 
\multirow{3}{*}{Concepts} 
& \multirow{3}{*}{
\begin{tabular}[c]{@{}l@{}} Various fairness concepts exhibit \\ both distinctions and \\ interconnected characteristics. \end{tabular}}  
& \multirow{3}{*}{
\begin{tabular}[c]{@{}l@{}} Create customized fairness \\ concepts for diverse \\recommendation scenarios. \end{tabular}}   
& \multirow{3}{*}{
\begin{tabular}[c]{@{}l@{}} Standardize industry concepts \\while promoting fairness \\ in diverse scenarios. \end{tabular}}

\\
\\
\\

\midrule 
\multirow{3}{*}{Frameworks} 
& \multirow{3}{*}{
\begin{tabular}[c]{@{}l@{}} There isn't a one-size-fits-all framework \\ for addressing fairness concerns. \end{tabular}}  
& \multirow{3}{*}{
\begin{tabular}[c]{@{}l@{}} Apply suitable fairness \\ methods to specific \\ fairness issues. \end{tabular}}   
& \multirow{3}{*}{
\begin{tabular}[c]{@{}l@{}} Ensuring the most effective \\ solutions are applied. \end{tabular}}

\\
\\
\\
\midrule

\multirow{3}{*}{Trade-off} 
& \multirow{3}{*}{
\begin{tabular}[c]{@{}l@{}} Balancing fairness can sometimes \\ affect accuracy, causing imperfect \\ outcomes. \end{tabular}}  
& \multirow{3}{*}{
\begin{tabular}[c]{@{}l@{}} Finding a balance \\ between fairness and \\ recommendation performance. \end{tabular}}   
& \multirow{3}{*}{
\begin{tabular}[c]{@{}l@{}}Ensuring optimal results \\ while promoting equitable \\treatment for all users.\end{tabular}}

\\
\\
\\

\midrule

\multirow{3}{*}{Trustworthiness} 
& \multirow{3}{*}{
\begin{tabular}[c]{@{}l@{}} Fairness-trustworthiness \\ interactions are underexplored \\ in current researches. \end{tabular}}  
& \multirow{3}{*}{
\begin{tabular}[c]{@{}l@{}} Explore the interaction rules \\ between fairness and \\ trustworthiness. \end{tabular}}   
& \multirow{3}{*}{
\begin{tabular}[c]{@{}l@{}}Foster trustworthy properties \\ in fairness RSs to \\ increase trustworthiness.\end{tabular}}

\\
\\
\\

 \bottomrule
\end{tabular}
}

\label{Futuretable}

\vspace{-0.5cm}
\end{table*}

\subsection{Connections with Others}
Discrimination occurs in an untrustworthy recommender system. Mansoury et al. \cite{DBLP:conf/flairs/MansouryASDPM20} propose three different user profile features, and analyze the possible connection between these features and the different behaviors of the recommender system for different genders. They introduce the unfairness of the recommender system caused by gender discrimination, and find that women get less accurate recommendations than males based on their experiments. This phenomenon indicates that the recommendation algorithm is unfair to different genders.

An ethical recommendation system cannot discriminate against vulnerable groups. In addition to protecting user privacy from the standpoint of ethical and moral norms, we also need to consider the requirements of relatively underprivileged groups.
A number of measures are suggested by Leonhardt et al. \cite{DBLP:conf/www/LeonhardtAK18}  as ways to quantify the influence of fairness-aware pre-processing techniques on user prejudice. 
A re-ranking method is developed by Gomez et al. \cite{Mooc} in order to reduce bias caused by the discrimination against teachers' geographic location.
By using black feminist and critical race theory, Schelenz \cite{DBLP:conf/um/Schelenz21} attempts to lessen the unfairness of the user's political and social environment.

In addition to the problem of discrimination, contemporary recommender systems use big data to conduct in-depth and detailed mining of users' historical behaviors, personal characteristics, and other data. They take the learned user's preference as the main standard and provide precise "Act According to Actual Circumstances" recommendations for the user. While these recommender systems excessively collect users' private data, they limit the possibility of ordinary people exploring a variety of new fields.

Another form of fairness that some studies suggest is diversity \cite{DBLP:conf/um/Schelenz21}. Diversity can alleviate the ethical issues caused by the precise recommendation of recommender systems. 
Mansoury et al. \cite{DBLP:journals/tois/MansouryAPMB22} solve the problem of unfairness by transforming it into a maximum flow problem, which improves the overall diversity and fair distribution of recommended items.
Sacharidis et al. \cite{DBLP:conf/um/SacharidisMW20} propose a regularization for social recommendations that allows friends to be similar. However, within a community, it generally forces members to be more diverse, which results in fairer recommendations.

\section{Future Directions}
\label{Future}

\begin{table*}[ht!]
\renewcommand\arraystretch{1.2}
\centering
\caption{A comprehensive overview of recommender systems that are both fairness-aware and incorporate trustworthiness features. The horizontal axis represents the recommendation methods promoting fairness, while the vertical axis corresponds to the trustworthiness features. The blank areas indicate an absence of related research in those particular domains. These unexplored areas also present opportunities for future research directions.
}
\label{fairtrust}
\begin{threeparttable}
\begin{tabular}{@{}lp{1.85cm}p{1.85cm}p{1.85cm}p{1.85cm}p{1.85cm}p{1.7cm}p{1.85cm}@{}}
\cline{2-8}
\multicolumn{1}{l|}{\begin{tabular}[c]{@{}l@{}}Explain\\ -ability\end{tabular}} & \multicolumn{1}{l|}{}                                                                           & \multicolumn{1}{l|}{}                                   & \multicolumn{1}{l|}{}                                      & \multicolumn{1}{l|}{\cite{DBLP:journals/corr/abs-2006-16977}, \cite{DBLP:conf/sigir/GeTZXL0FGLZ22}} & \multicolumn{1}{l|}{}                                            & \multicolumn{1}{l|}{\cite{fu2020fairness}} & \multicolumn{1}{l|}{}                                                      \\ \cline{2-8} 
\multicolumn{1}{l|}{\begin{tabular}[c]{@{}l@{}}Robust\\ -ness\end{tabular}}     & \multicolumn{1}{l|}{\cite{DBLP:conf/sacmat/Fang0MS22}, \cite{DBLP:conf/wsdm/RastegarpanahGC19}} & \multicolumn{1}{l|}{\cite{zhu2022cali3f}}               & \multicolumn{1}{l|}{}                                      & \multicolumn{1}{l|}{\cite{dokoupil2022robustness}}                                                  & \multicolumn{1}{l|}{}                                            & \multicolumn{1}{l|}{}                      & \multicolumn{1}{l|}{}                                                      \\ \cline{2-8} 
\multicolumn{1}{l|}{Privacy}                                                    & \multicolumn{1}{l|}{}                                                                           & \multicolumn{1}{l|}{}                                   & \multicolumn{1}{l|}{\cite{wu2021learning}}                 & \multicolumn{1}{l|}{\cite{wei2022comprehensive}}                                                    & \multicolumn{1}{l|}{}                                            & \multicolumn{1}{l|}{}                      & \multicolumn{1}{l|}{\cite{do2022online}}                                   \\ \cline{2-8} 
\multicolumn{1}{l|}{Others}                                                     & \multicolumn{1}{l|}{}                                                                           & \multicolumn{1}{l|}{\cite{DBLP:conf/um/SacharidisMW20}} & \multicolumn{1}{l|}{}                                      & \multicolumn{1}{l|}{}                                                                               & \multicolumn{1}{l|}{}                                            & \multicolumn{1}{l|}{}                      & \multicolumn{1}{l|}{\cite{Mooc}, \cite{DBLP:journals/tois/MansouryAPMB22}} \\ \cline{2-8} 
                                                                                & \begin{tabular}[c]{@{}l@{}}Pre-processing\\Methods\end{tabular}                                & \begin{tabular}[c]{@{}l@{}}Regularization \\  \end{tabular}
                                                                                                                          & \begin{tabular}[c]{@{}l@{}}Causal\\ Inference\end{tabular} & \begin{tabular}[c]{@{}l@{}}Adversarial\\ Learning\end{tabular}                                      & \begin{tabular}[c]{@{}l@{}}Reinforcement\\ Learning\end{tabular} & 
                                                                                                                          
                                                                                               \begin{tabular}[c]{@{}l@{}}Ranking \\  \end{tabular}                                                               & \begin{tabular}[c]{@{}l@{}}Post-processing \\ Methods\end{tabular}         \\ 
\end{tabular}
\end{threeparttable}
\end{table*}

In recent years, attention from the academy and industry communities has been paid to improving fair recommender systems. However, thoroughly building fair recommender systems that can be trusted still faces the following challenges. Table \ref{Futuretable} outlines current works' shortcomings, unresolved issues, and potential benefits of resolving them.

\noindent\textbf{Concepts for fairness.} 
In different recommendation scenarios, the fairness goals that people pursue are also different. For example, in a recommender system with multiple stakeholders, the goal of fairness is to balance the interests of multiple stakeholders. However, the goal of fairness in a time-aware job recommender system is to balance exposure frequency with old and new job information \cite{10.1145/3357384.3358131}.
The concepts of fairness in different scenarios have both mutually inclusive and different parts. For example, both long-term fairness and dynamic fairness are fairness affected by time. A statically fair recommender system may include individual fairness and group fairness. Therefore, it is difficult to have a unified and accurate concept to define the fairness of recommender systems. Obtaining a common concept for different definitions of fairness is an important challenge. In addition, there may be some different fairness issues in a recommendation scenario, and these fairness issues may be conflicting. A potentially promising approach is to consider the prioritization of fairness issues in a recommender system. Few works consider the importance of fairness issues. This will also be an important challenge in the future.

\noindent\textbf{General frameworks for fairness.}
The application scenarios of fairness-aware recommender systems are broad, including education, society, health care, et al. However, each recommendation scenario has different fairness-aware recommendation methods, such as regularization-based methods and re-ranking methods. At present, there is no work to analyze which type of fairness method is applicable to a certain type of fairness issue from the perspective of recommendation scenarios. In addition, due to the diversity of fairness issues, building a unified recommendation framework to solve all fairness issues can simplify the analysis of various fairness issues and also can be quickly applied to new scenarios and unknown fairness issues. There is no work yet to solve the above problem. We look forward to building a general recommendation framework to address different fairness issues in the future.

\noindent \textbf{Trade-off between fairness and performance of RSs.}
As a means of ensuring the fairness of the recommender system, a system needs to reduce the bias in the recommendation output. However, it is challenging to combine fairness with accuracy in recommender systems, mainly because the goals of fairness and accuracy are inconsistent and the trade-off between them is substantial. In recommender systems, accuracy is determined by the system's capacity to accurately anticipate and meet the needs and interests of users. Taking the example of a multi-stakeholder recommender system (such as suppliers and consumers), users want the recommendation system to recommend products that meet their preferences, and suppliers want to make the items they provide as fair as possible to recommend to users. If the fairness of item exposure is maintained, it may lead to a decrease in the accuracy of recommendations. Therefore, controlling the trade-off between accuracy and fairness becomes critical.

\noindent\textbf{Mutual promotion with trustworthy properties.} At present, building a trustworthy recommendation system is the development trend of artificial intelligence. Credibility includes multiple properties, and fairness is one of them. Most of the current work focuses on the association between fairness and trustworthiness, and does not study the law of mutual influence between these properties. For example, the work of Fu et al. \cite{fu2020fairness} introduces achieving fairness on an explainable recommender system. This type of work mainly addresses the issue of fairness without improving explainability. In other words, these approaches to fairness do not promote explainability. We hope that future works can focus on the intrinsic interaction between fairness and other trustworthy properties in recommender systems. In addition to explainability, there is also little work that considers the positive and negative effects between robustness and fairness.

Most of the current fairness work coexists with only one fairness property, and no work comprehensively considers all fairness properties. Another challenge in the future is how to integrate fairness and other credible properties in a recommender system, establish internal correlations between fairness and properties, reduce conflicts between properties, and build a credible recommender system.
Table \ref{fairtrust} shows the current state of research on integrating fairness methods and trustworthiness properties. It's evident that numerous areas have yet to be explored, such as the application of reinforcement learning methodologies to develop explainability and fairness recommender systems. There remains significant research potential in exploring the integration of trustworthiness and fairness.

\section{Conclusion}
\label{Conclusion}
Recommender systems provide basic artificial intelligence services to facilitate our daily lives.
To ensure that fairness-aware recommender systems are crucial for people to enjoy trustworthy recommendations, This survey reviews current efforts on fair recommender systems with the aim of facilitating their implementation and future research.
Firstly, we establish a foundation by introducing fundamental concepts related to recommender systems and fairness. This introduction will later aid us in defining fairness in various contexts. Then, we introduce different types of biases that cause unfairness at each stage of the recommender system's lifecycle. Finally, we introduce the manifestations of fairness in different recommendation scenarios, which paves the way for the introduction of subsequent methods.
For fairness-aware recommendation methods, we introduce three processing stages: pre-processing, in-processing, and post-processing. We summarize the categories of methods and divide them in detail in each processing stage. For the evaluation of the fairness-aware recommender system, we introduce the data sets suitable for different recommendation scenarios and the fairness measurement metrics.
For the application of fairness-aware recommender systems in the real world, we describe how fairness-aware recommender systems maintain fairness in areas closely related to people's lives, such as e-commerce, education, and et al.
Regarding the promotion of human trust in fairness-aware recommender systems, we start with several trustworthy properties and introduce the correlation between fairness and these trustworthy properties. We conclude this survey and discuss the future directions of fairness-aware recommender systems from several novel perspectives, including the development direction of fair recommender systems, as well as the development of fair and trustworthy recommender systems.

\balance
\bibliographystyle{cas-model2-names}

\bibliography{sample-base}



\end{document}